\documentclass[useAMS,usenatbib]{mn2e}

\usepackage{graphicx}
\usepackage{amsmath}
\usepackage{amssymb}
\usepackage{fixltx2e}

\usepackage{color}

\def\electron{e}
\def\rin{r_{\rm in}}
\def\rout{r_{\rm out}}
\def\tin{T_{\rm in}}
\def\bin{B_{\rm in}}
\newcommand{\kom}[1]{\textcolor{black}{#1}}

\title[Gamma-rays from cascades in microquasars]
{Gamma-rays from anisotropic IC $e^\pm$ pair cascades
in microquasars: Application to Cyg~X-3}

\author[J. Sitarek~and~W. Bednarek]{J. Sitarek and W. Bednarek \\
Department of Astrophysics, University of \L\'od\'z, PL-90236 \L \'od\'z, Poland; jusi@kfd2.phys.uni.lodz.pl; bednar@astro.phys.uni.lodz.pl}

\begin{document}

\date{Accepted . Received ; in original form }

\pagerange{\pageref{firstpage}--\pageref{lastpage}} \pubyear{2010}

\maketitle

\label{firstpage}

\begin{abstract}
Microquasars have been expected to emit high energy $\gamma$-rays
due to their general similarities to the $\gamma$-ray emitting blazars (evidences of relativistic jets, non-thermal radio to X-ray emission).
In fact, the first source of this type, Cyg~X-3, has been recently unambiguously discovered by the satellite telescopes. 
We study the features of the $\gamma$-ray radiation produced in these sources by relativistic electrons, accelerated in the inner part of the jet.
The electrons initiate an Inverse Compton $e^\pm$ pair cascade in the radiation field of the accretion disk. 
Due to the anisotropy of the accretion disk radiation field, the spectra of $\gamma$-rays show strong dependence on the observation angle, the location of the emission region within the jet and the details of the acceleration process. As an example, we confront our model with the observations of the microquasar Cyg~X-3, which has been recently reported as a transient GeV $\gamma$-ray source by the Agile and the Fermi Observatories. Satisfactory description of the $\gamma$-ray spectra observed from Cyg~X-3 are obtained in the case of the injection of electrons in the inner part of the jet (located within 300 inner disk radius from the jet base) provided that the observer is located at relatively small angle to the jet axis.  
\end{abstract}

\section{Introduction}\label{sec:intro}

Microquasars are binary systems which show evidences of relativistic collimated outflows resembling those observed in active galactic nuclei
\citep[e.g. ][]{mr94, tin95, hjel95}.
They are interpreted in the model of a jet produced in the inner part of the accretion disk around solar mass black hole which accretes the matter from a close companion star \citep[e.g. ][]{lb96, aa99, gak02, rkm02, brp04}. 
Up to now, only one source of this type, i.e. \mbox{Cyg~X-3}, has been clearly detected in the GeV $\gamma$-rays by the Agile and the Fermi observatories \citep{tetal09, ab09a}. At the TeV energies only the upper limits are available in spite of intensive monitoring \citep{al10}. 
There are also some evidences of the sporadic GeV-TeV $\gamma$-ray emission from another source of this type, i.e. \mbox{Cyg~X-1} \citep{al07, sab10}. Other microquasars, such as GRS 1915+105 and SS433, have been also observed in the TeV energies by the Cherenkov telescopes but without positive result \citep[see e.g. ][]{aha05, hay09, ac09, sea09, al11}. 
The GeV-TeV emission has been also detected from a few other massive binary systems, i.e. LS5039 and \mbox{LSI 61 303}. However it seems at present that they do not  belong to the class of microquasars as it was originally suspected (see e.g. \citealp{du06}).

It is expected that the $\gamma$-ray emission from  microquasars is produced by relativistic electrons which are accelerated in the mildly relativistic jet. 
These electrons scatter the soft radiation from the accretion disk \citep[e.g.][]{cer11}, or the companion star \citep[e.g.][]{bed10, dch10}. 
If hadrons are as well accelerated in jets of microquasars, then their $\gamma$-ray radiation should be accompanied by the neutrino fluxes potentially observable by the large scale neutrino telescopes \citep[e.g.][]{lw01, dgwl02, rtkm03, bed05}. 

We propose that the $\gamma$-ray emission from microquasars is mainly produced within the inner part of the jet 
in the anisotropic Inverse Compton $e^\pm$ pair cascade initiated by relativistic electrons in the radiation field of the accretion disk. 
If the relativistic electrons are injected close to the disk, then the disk radiation dominates over the radiation from the companion star. 
In the example calculations we apply the parameters of the Cyg~X-3 binary system which contain very hot WR type star and an object on a very compact orbit. 
For such luminous stars in compact binaries as Cyg~X-3, the TeV $\gamma$-rays produced within the binary system should be efficiently absorbed in the radiation field of the companion star \citep[e.g.][]{ap84, ps87, mkt93}. 
If electrons are accelerated close to the surface of the accretion disk, then $\gamma$-rays can be at first absorbed in the X-ray radiation coming from the disk itself (see calculations in the case of X-ray binaries in e.g. \citealp{car92, bed93, cer11} and in the case of active galaxies in \citealp{bk95, sb08}). The $\gamma$-ray spectrum emerging from the radiation of the accretion disk is modified in such a way that its farther absorption in the stellar radiation becomes negligible in most cases.
Therefore, $\gamma$-ray spectra escaping from the compact, massive binary systems, such as Cyg~X-3, should form in the cascade process initiated by relativistic electrons in the anisotropic radiation field of the accretion disk as recently considered in the case of active galaxies
\citep{sb10a,sb10b}.
In this paper we calculate for the first time the angle dependent $\gamma$-ray spectra produced in the anisotropic IC $e^\pm$ pair cascade initiated by relativistic electrons in jets of  microquasars. 
As an example, we confront these spectra with the GeV $\gamma$-ray observations of Cyg~X-3 by the Fermi-LAT telescope.

\section{A model for gamma-ray production in IC $\electron^\pm$ pair cascade}

\begin{figure}
\includegraphics[width=0.47\textwidth]{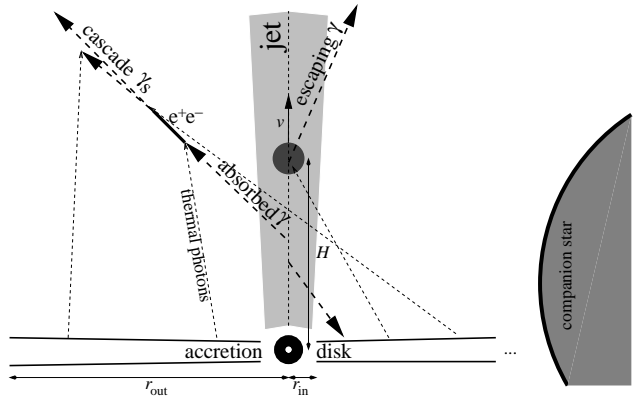}
\caption{
Schematic representation of the scenario considered in the paper.
IC $e^\pm$ pair cascades are initiated by relativistic electrons contained within the emission region (the blob) moving along the jet axis. \
Electrons scatter in the IC process the anisotropic radiation coming from the accretion disk formed by the matter accreting from the companion star. 
$\gamma$-rays (the thick dashed arrows) are produced in the blob (the dark circle) moving along the jet (the light shaded region) via the IC scattering of the thermal disk photons (the thin dashed arrow). 
Some of them escape to the observer but others either hit the accretion disk surface or they are absorbed in the thermal disk radiation and produce $e^+e^-$ pairs (the thick solid line). 
These $e^\pm$ pairs can generate the second generation of $\gamma$-rays. The companion star is shown as the big circle on the right. 
The accretion disk spreads throughout the range of distances $\rin - \rout$ and emits black body radiation with the temperature profile $T \sim r^{-3/4}$.
}
\label{fig_scenerio}
\end{figure}

We adopt the classical geometry of a microquasar in which a jet is launched from the inner part of the accretion disk, perpendicularly to its plane. The disk is formed by the matter which is supplied by the companion star (see Fig.~\ref{fig_scenerio}). 
\kom{This matter is supplied to the compact object in the form of a fast wind. 
However, this wind have significant angular momentum in the rest frame of the compact object due to its fast orbital motion around the massive star. 
Therefore, the formation of the relatively thin accretion disk in the inner part of the accretion flow is very likely.}
The standard accretion disk model is applied in which the disk surface temperature is described by the power law profile: $T(r)=\tin(r/\rin)^{-3/4}$, where $\tin$ is the temperature at the inner radius of the disk, $\rin$ (see \citealp{ss73}).
For the example calculations presented below we assume typical values for the accretion disk, $\tin=5\cdot10^6$~K and $\rin=10^7$~cm. 
For these parameters the disk luminosity is equal to $L_{\rm d} = 4\pi \rin^2\sigma_{\rm SB}\tin^4\approx  4.5\times 10^{37}$ erg s$^{-1}$, where $\sigma_{\rm SB}$ is the Stefan-Boltzmann constant, which is typical in the case of X-ray binary systems. 
We assume that the jet has a conical structure with the opening semi-angle of the order of $\alpha=5^\circ$.
Assuming that the total flux of the magnetic field through the cross section of the jet is conserved, we estimate the magnetic field strength in the jet at the height $h$ above the accretion disk on:
\begin{equation}
B(h) = \bin\rin^2/(\rin+h\sin\alpha)^2,
\end{equation}
where  $\bin$ is the strength of the magnetic field at the height $\rin$ above the disk.
We relate the free parameter $\bin$ to the inner disk temperature by assuming some level of the equipartition, $\eta$, between the energy density of the magnetic field and the energy density of the isotropic black body radiation with the fixed temperature of $\tin$: $\bin\approx 4\cdot10^5 \sqrt{\eta} (\tin/\mathrm{10^6K})^2$~G.
Note that due to the assumed temperature profile and the geometry of the radiation field close to the disk, even for $\eta=1$ the magnetic field energy density will dominate over the energy density of the radiation field.
For example, at the height of $1\,\rin$, for $\eta=1$ the magnetic field energy density is larger by one order of magnitude.

\begin{figure}
\includegraphics[width=0.49\textwidth]{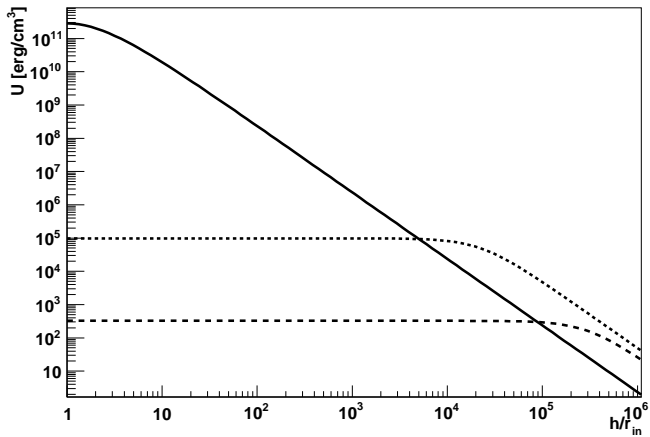}
\caption{
Energy density of the radiation of the accretion disk, with the temperature $\tin=5\cdot10^{6}$~K at its inner radius $\rin=10^{7}$~cm, as the function of the height above the disk (the solid curve). 
The energy density of the radiation from the companion stars: in Cyg~X-1 binary system (the surface temperature of the companion star $T_\star=3.2\cdot 10^4$~K, the radius $R_\star=1.2\cdot 10^{12}$~cm, and the separation $a=2.47\,R_\star$, the dashed curve) and Cyg~X-3 ($T_\star=9\cdot 10^4$~K, $R_\star=2\cdot 10^{11}$~cm, and $a=2.25\,R_\star$, the dotted curve).
}
\label{fig_energydensity}
\end{figure}

In order to find out which radiation field is mostly relevant in the $\gamma$-ray production in different parts of the jet, we compare the energy densities of the radiation fields from the accretion disk and from the companion star.
We assume typical parameters of the companion stars and of the binary systems as observed in two well known binaries, i.e. Cyg~X-1 and Cyg~X-3 (see Fig.~\ref{fig_energydensity}). 
It is clear that the energy density of the accretion disk dominates up to the height comparable with the separation, $a$, of the components of the binary system, i.e. $H\sim 10^4 - 10^5 \rin\sim a$. 
Therefore, we conclude that, except of the most outer parts of the jet, it is likely that particles accelerated in the jet will predominantly interact with the radiation field of the accretion disk in these two binaries. 
Note, moreover that the surface temperature of the companion star is about two orders of magnitude lower than the temperature at the inner disk radius. 
Because of that, we can safely neglect the effects of farther absorption and cascading of the $\gamma$-rays, produced in the considered model, in the radiation field of the companion star.  

We consider the scenario in which electrons are accelerated in jets of microquasars in the classical shock acceleration mechanism. 
The electrons obtain the power law spectrum and they are distributed isotropically in the reference frame of the emission region (the blob).
The blob moves relativistically along the jet with the specific velocity $v_{\rm b}$, resulting in an anisotropic distribution of the electrons in the disk frame of reference. 
The electrons interact with the radiation of the disk producing the first generation of the $\gamma$-rays in the Inverse Compton scattering process. 
While the electrons are confined in the jet by the magnetic fields and move with the bulk velocity of the blob, the $\gamma$-rays produced by them escape from the jet into the volume above the accretion disk.
The primary $\gamma$-rays propagate in the dense radiation field of the accretion disk. 
Significant part of them is absorbed, producing the first generation of $e^\pm$ pairs. 
As the result of the IC and $\gamma$-$\gamma$ absorption processes, the IC $e^\pm$ pair cascade develops in the whole volume above the accretion disk \citep[see also][]{sb10a, sb10b}. 
In the cascade process considered in this paper we neglect the effects of bending of the trajectories of the secondary leptons by the local magnetic field which could be present above the disk. 
In fact, these effects can be neglected in the case of a dipole magnetic field above the accretion disk.
Such field drops fast with the distance from the centre of the disk (proportionally to $\propto R^{-3}$) and is important only very close to the disk (see \citealp{sb10a}). 
Since the disk radiation field is highly anisotropic, the $\gamma$-ray spectra produced in such a cascade process strongly depend on the location of the observer in respect to the axis of the accretion disk. 
In the subsections below we consider the input parameters for such a model and show the conditions for which the $\gamma$-rays can both be efficiently produced and escape to the observer.

\subsection{Acceleration of electrons in the jet}\label{sec:accel}

\begin{figure*}
\includegraphics[width=0.3275\textwidth]{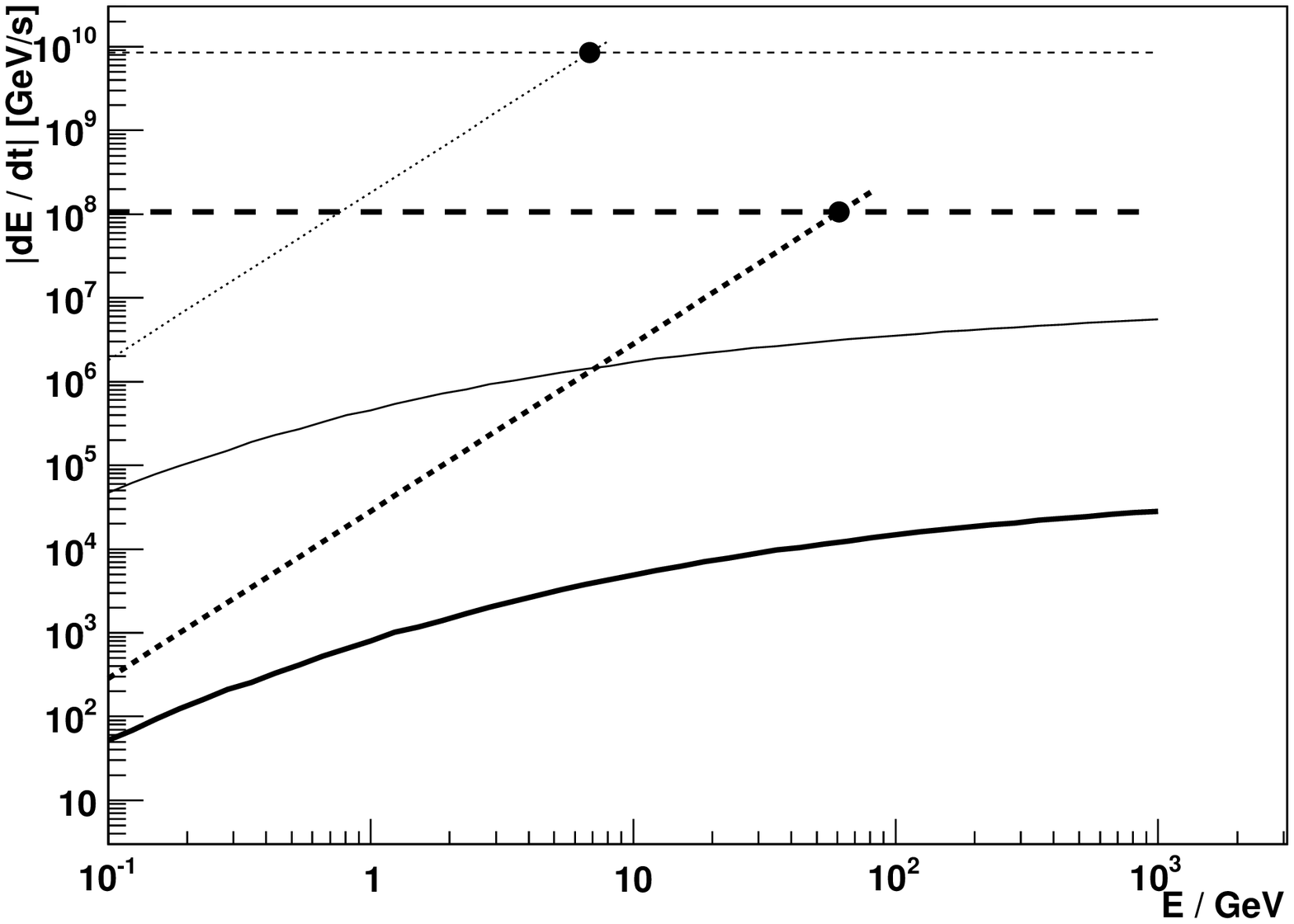}
\includegraphics[width=0.3275\textwidth]{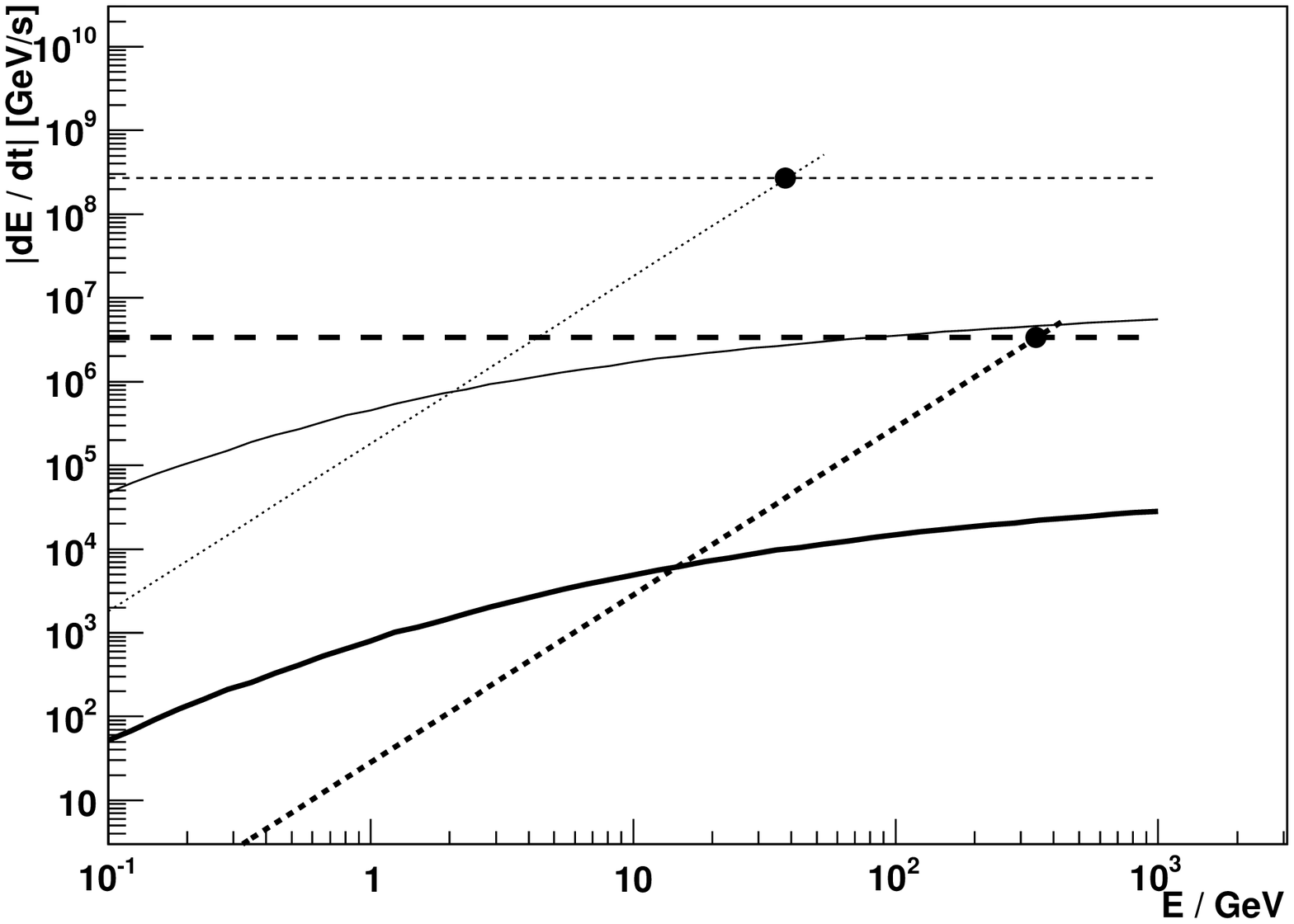}
\includegraphics[width=0.3275\textwidth]{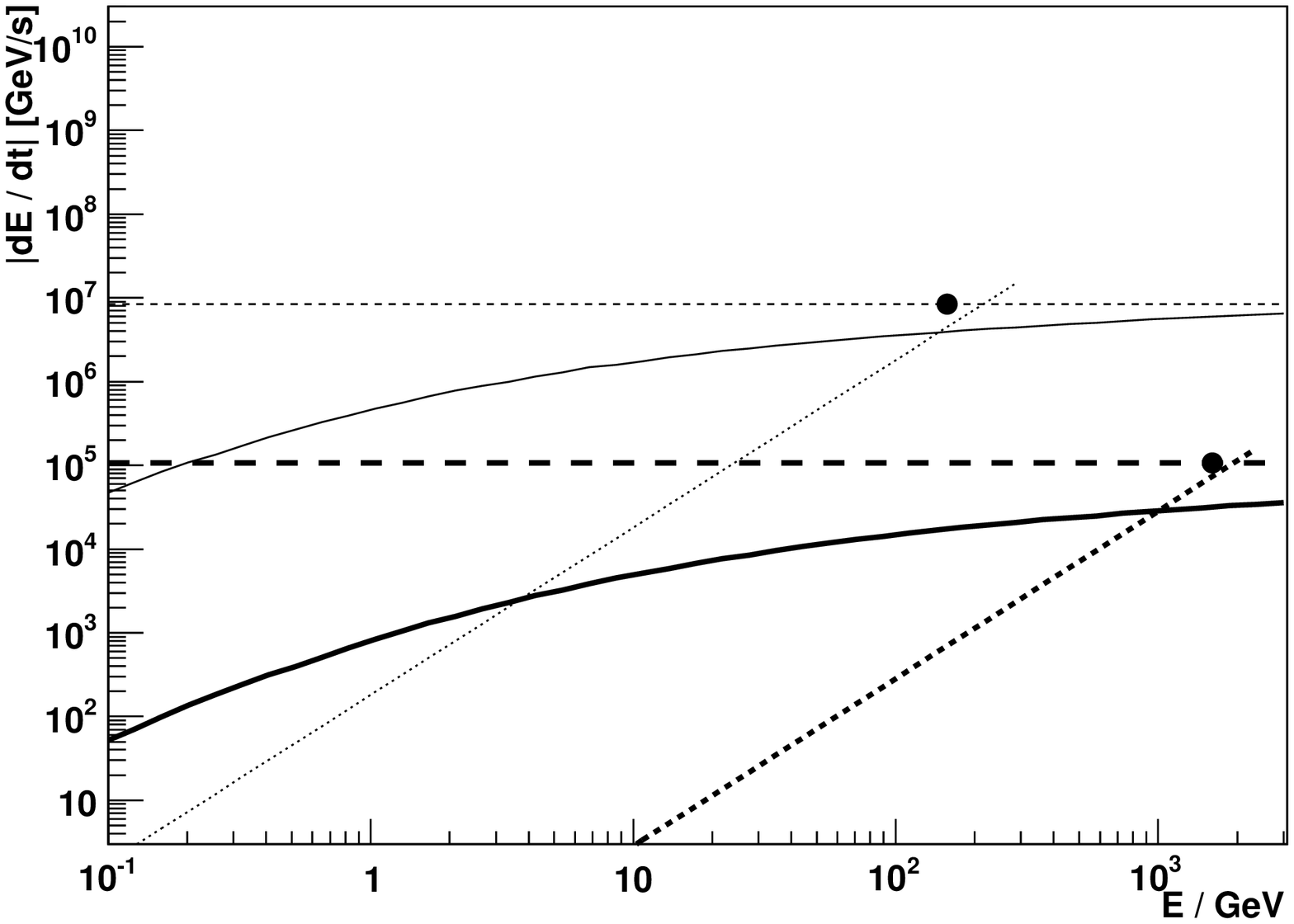} \\
\includegraphics[width=0.3275\textwidth]{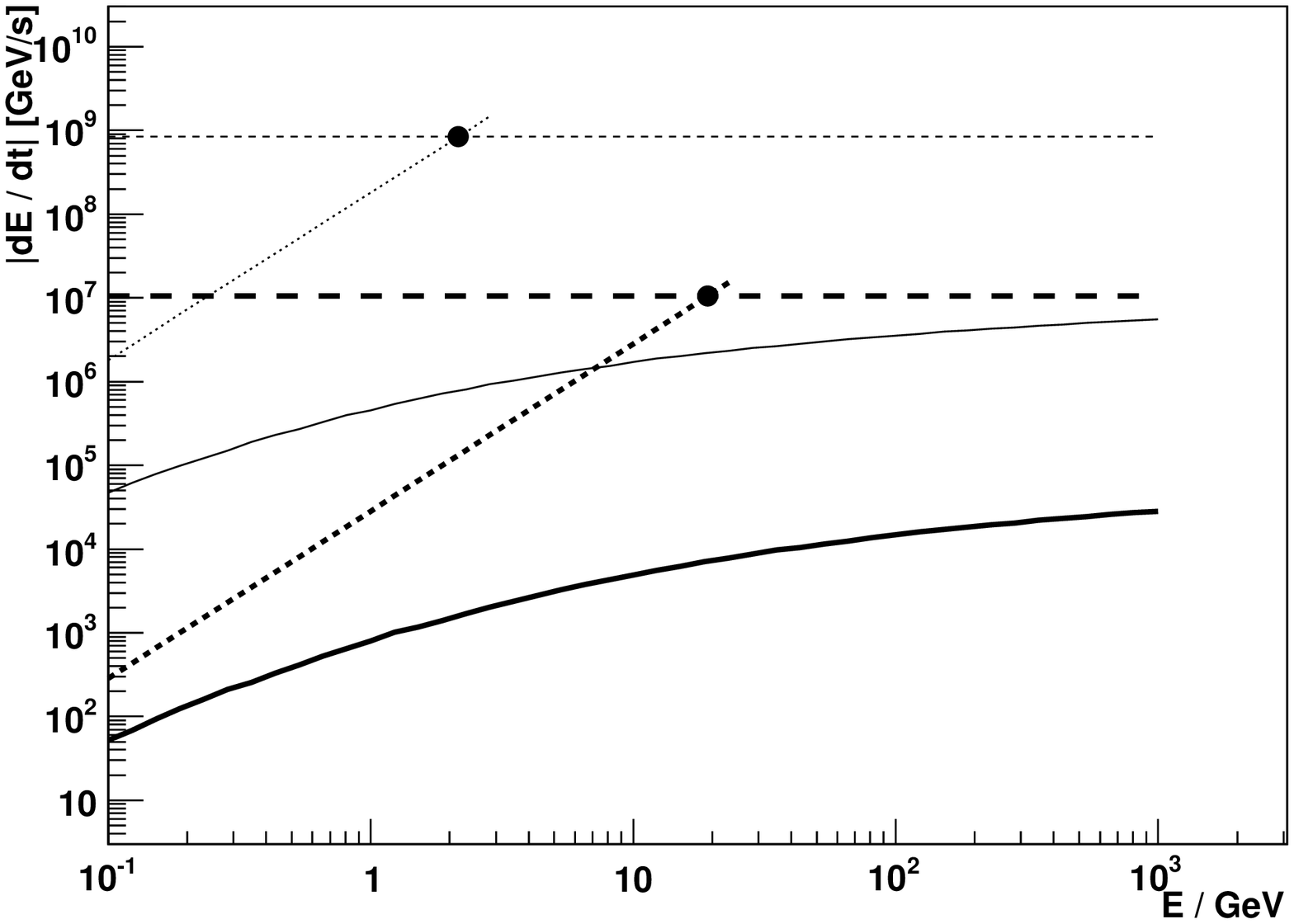}
\includegraphics[width=0.3275\textwidth]{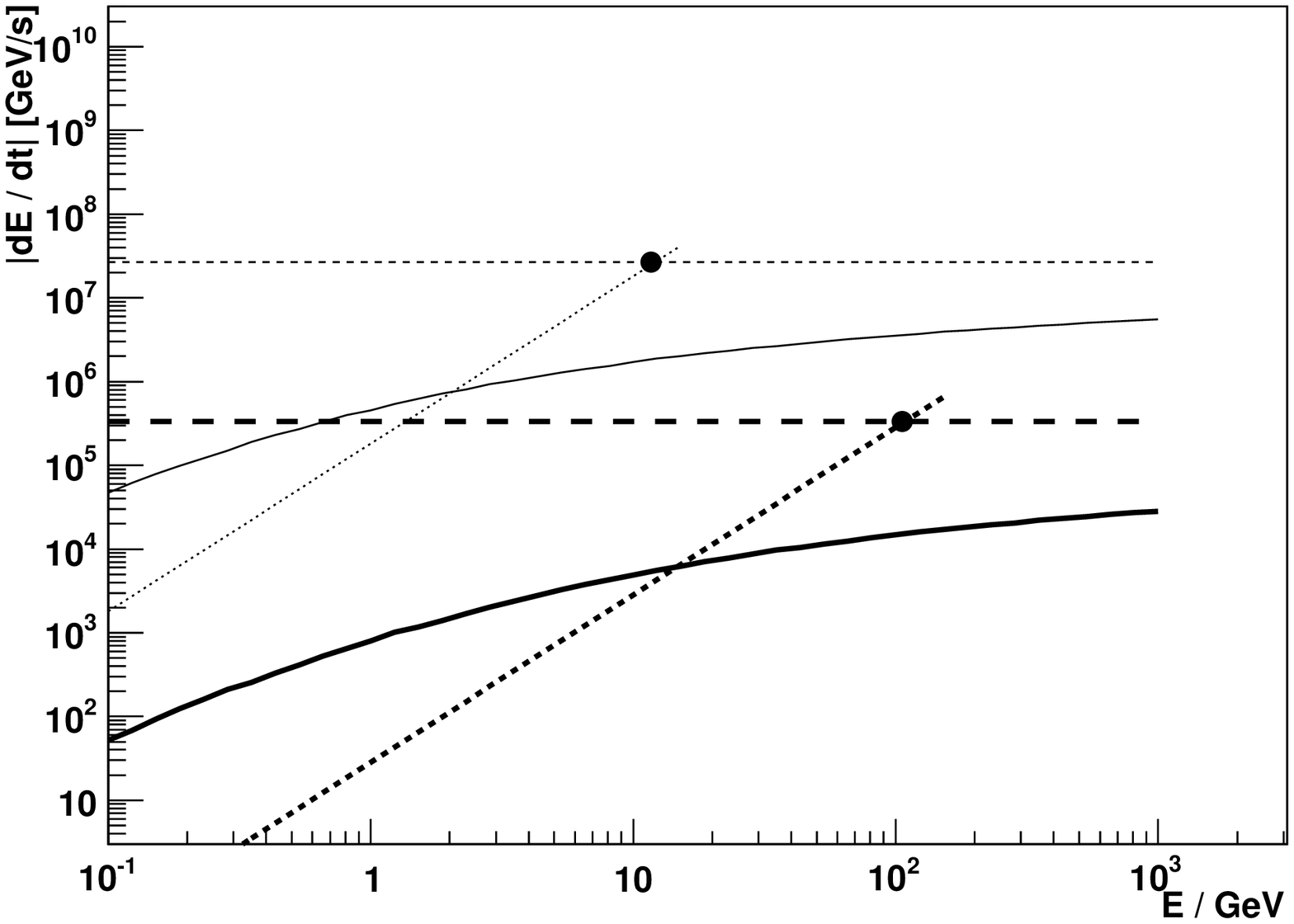}
\includegraphics[width=0.3275\textwidth]{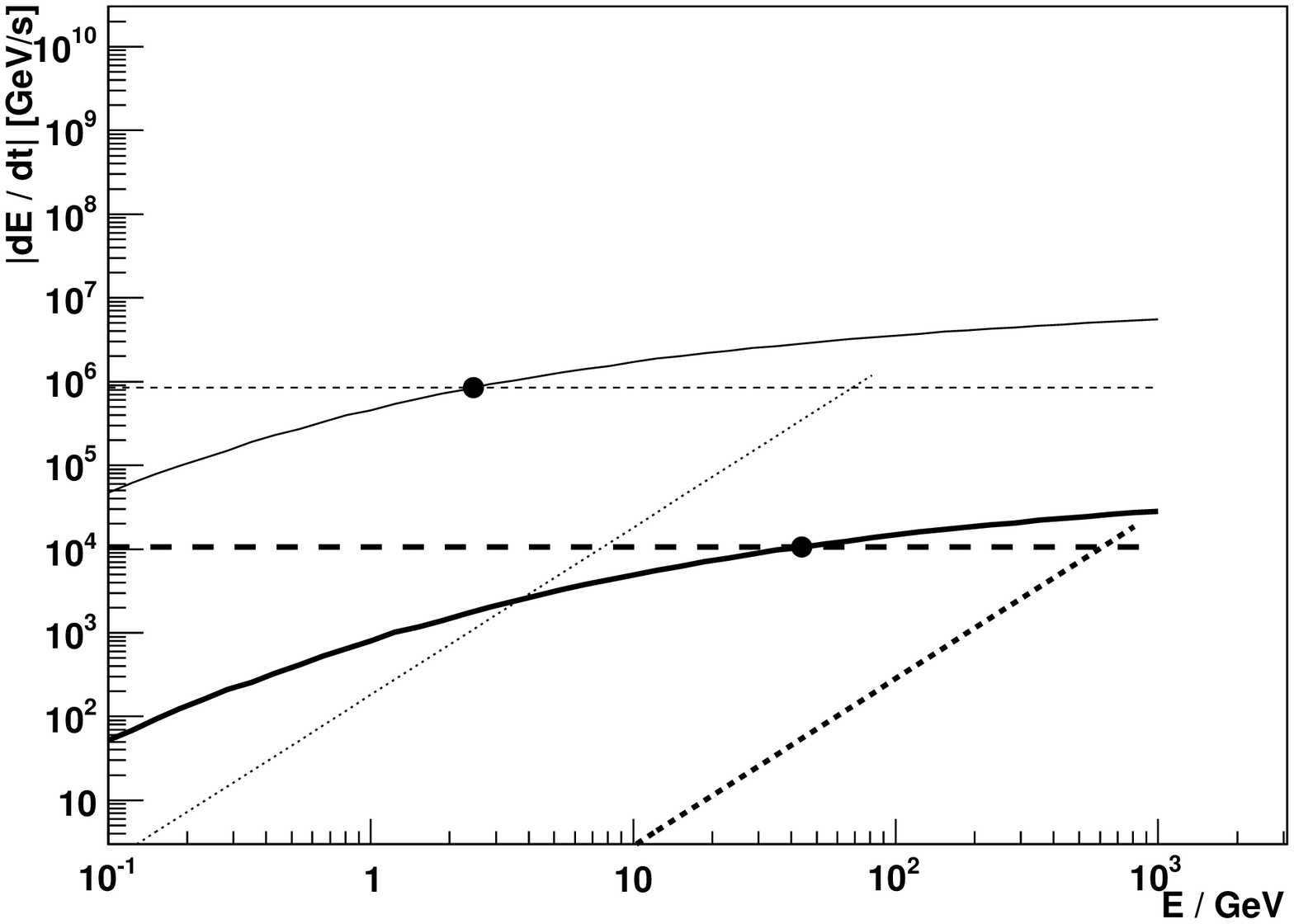} \\
\caption{
Energy gains (the horizontal dashed lines), synchrotron (the dotted lines) and IC (the solid curves) losses of the accelerated electrons at the height $H=1\,\rin$ (the thin lines) or $H=100\,\rin$ (the thick lines) above the accretion disk.
The equipartition parameter between the magnetic and the radiation fields energy density at the inner radius (see text for details) is assumed to be $\eta=1$ (the left panels), $10^{-3}$ (middle), and $10^{-6}$ (right). 
The calculations are shown for different acceleration efficiencies of electrons: $\xi=0.1$ (the top panels) and $0.01$ (bottom). 
The disk temperature at the disk inner radius, $\rin=10^7$~cm, is $\tin=5\cdot 10^{6}$~K. The circles mark the intersections at which the energy gains balances the energy losses of electrons.} 
\label{fig_straty}
\end{figure*}
The rate of the energy gain  of an electron with the energy $E$ due to the acceleration process in the jet can be generally parametrized as: 
\begin{equation}
(dE/dt)_{\rm acc} = \xi  c E/R_{\rm L}\approx 10^{13}\xi B\ ~~~{\rm eV~s^{-1}},\label{eq2}
\end{equation}
\noindent
where $\xi$ is the so-called acceleration coefficient, $R_{\rm L}$ is the Larmor radius of electrons, and $B$ is the magnetic field strength (in Gauss) at the acceleration region. 
We estimate the acceleration coefficient on $\xi\sim (v/c)^2$ in the case of the acceleration process occurring in the turbulent medium of the jet and/or multiple shocks, where $v$ is the velocity of the shock and $c$ is the speed of light. 

The primary electrons accelerated in the jet lose their energy due to both the synchrotron and the IC cooling. 
Which of those dominates at a given location in the jet depends on the assumed parameters of the model, especially on the level of the equipartition between the magnetic field and the disk radiation field energy densities, $\eta$. 
The synchrotron energy loss rate can be calculated from the formula: 
\begin{equation}
(dE/dt)_{\rm syn} = {\frac{4}{3}}c\sigma_{\rm T}U_{\rm B}\gamma_{\rm e}^2\approx 8.6\times 10^{-4}B^2\gamma_{\rm e}^2~~~{\rm eV~s^{-1}},\label{eq3}
\end{equation}
where $\gamma_{\rm e}$ is the Lorentz factor of the electron.

The energy loss rate due to the IC scattering is more difficult to compute because of the strongly anisotropic radiation field and the complicated cross section. 
For the considered parameters, the electrons scatter the soft radiation mainly in the Klein-Nishina (KN) regime of the IC process.
Therefore, the synchrotron energy losses can dominate even for relatively small values of $\eta$. 
We calculate numerically the energy losses of electrons on these two processes as a function of their energy and the distance from the disk surface (applying the full KN cross section). 
The energy gain and loss rates of the electrons are compared in Fig.~\ref{fig_straty} for different sets of the parameters describing the content of the jet, the acceleration process, and the distance from the base of the jet.
Note that except for the case of a very weak magnetic field in the jet and a large distance from the base of the jet, the acceleration process of electrons is saturated by the synchrotron energy losses. 
However, just below this saturation energy the IC energy losses can be easily comparable or even much larger than the synchrotron energy losses.
Therefore, we conclude that the significant amount of energy accumulated in primary electrons should be converted to the $\gamma$-rays in the IC process of the disk radiation. 
Closer to the base of the jet, where the magnetic field is stronger (see Eq.~\ref{eq2} and Eq.~\ref{eq3}), the energy loses grow faster than the energy gains, resulting in lower values of the maximum energy of the accelerated electrons. 
At $H\sim \rin$, electrons can reach maximum energies only of the order of a few to a few tens GeV. However, already at the distance of $H\sim 10^2\,\rin$, electrons can be accelerated up to $\sim 10^2-10^3$ GeV. 
Therefore, in principle, $\gamma$-rays with energies clearly above the threshold of even present Cherenkov telescopes should be produced in such a model.

The maximum energies of the electrons, estimated above, are correct provided that their acceleration time is significantly shorter than the ballistic time of the plasma flow through the jet, $\tau_{\rm b}$, so the electrons can be considered as confined at a given location in the jet. 
Only in such a case, the electrons can reach large energies close to the accretion disk, the source of the strong radiation. 
The ballistic time scale for the jet plasma can be estimated from, 
\begin{equation}
\tau_{\rm b} = H/v_{\rm b}\approx 3.3\times 10^{-4}H_{\rm in}/\beta~~~{\rm s},
\end{equation}
\noindent
where $H_{\rm in}=H/\rin$ is the distance measured from the base of the jet in units of the inner radius of the disk, and $\beta = v_{\rm b}/c$.
Electrons are accelerated locally within the jet ($\tau_{\rm acc} < \tau_{\rm b}$) for the condition,
\begin{equation}
E_{\rm acc} < 3.3 H_{\rm in}\xi B/\beta~~~{\rm GeV}.
\end{equation}
\noindent
Let us introduce some extreme parameters for this last formula.
The magnetic field strength in the inner jet is of the order of $B_{\rm in}\approx 10^4$ G, even for very small $\eta$ (e.g. for $\eta = 10^{-6}$, and $H_{\rm in} = 1$. The typical values for the acceleration parameter in the relativistic shocks are $\xi \gtrsim 0.01$ and the velocities of the microquasars jets are estimated on $\beta =0.3$. Even for this limiting parameters, these critical energies of accelerated electrons are of the order of $E_{\rm acc}\approx  10^3$ GeV. 
We conclude that in the model considered by us, the electrons can be accelerated locally in the jet even up to TeV energies. 
Note that the velocity of the jet in Cyg~X-3 is measured in the range $0.3c-0.8c$ \citep{sp86, mio01, mar01}.
While, as the above calculation show, the acceleration in our model can be considered as local, the cooling process of electrons does not need to occur locally in the jet. 
Therefore, we include in our numerical code the effects of the change of location of blob, containing the primary electrons, in the jet during its propagation with the jet flow.

\subsection{Absorption of gamma-rays in the disk radiation}

\begin{figure}
\includegraphics[width=0.49\textwidth]{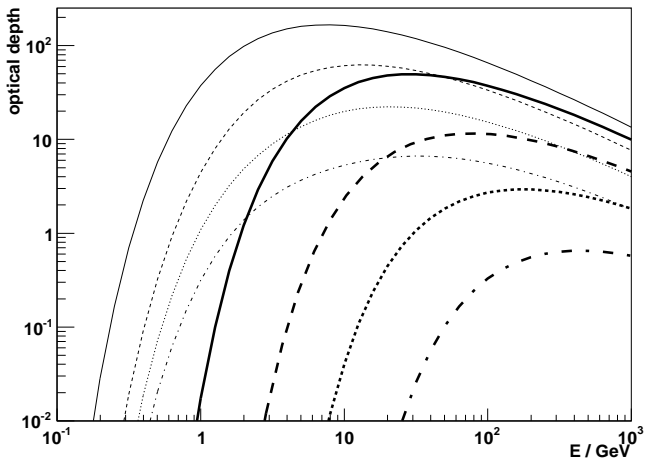}
\caption{
Optical depths of $\gamma$-rays in the radiation field of the accretion disk with the surface temperature defined by the Shakura \& Sunyaev profile, and the temperature $\tin=5\cdot10^{6}$~K at the inner radius of the disk $\rin=10^{7}$~cm as a function of their energies.
The optical depths for the $\gamma$-rays propagating along the jet axis ($\theta=0^\circ$) are marked by the thick curves and for $\theta=60^\circ$ by the thin curves. 
The different curve styles correspond to different injection height of $\gamma$-rays above the accretion disk: 
$H=3\,\rin$ (solid), $10\,\rin$ (dashed), $30\,\rin$ (dotted), and $100\,\rin$ (dot-dashed).
}
\label{fig_tau}
\end{figure}

As it was discussed in Sect.~\ref{sec:intro}, $\gamma$-ray photons propagating close to the accretion disk can be absorbed producing secondary $e^\pm$ pairs. 
The optical depths for $\gamma$-rays will strongly depend on their injection distance and the propagation angle, measured in respect to the axis of the disk (see e.g. \citealp{sb08, cer11}). 
As an example, we show the calculations of the optical depths \kom{as the function of the energy of the gamma-rays (calculated in the reference frame of the disk)} in the case of the assumed by us in the Sect.\ref{sec:accel} accretion disk parameters, selected distances from the base of the jet, and two injection angles of $\gamma$-rays (see Fig.~\ref{fig_tau}). In general, the optical depths can easily reach values up to $\sim 100$.
In contrary to the case of blazars, the typical accretion disks occurring in microquasars have much larger surface temperatures.
This results in an efficient absorption of the $\gamma$-rays already with energies of a few GeV, 
provided that they are already produced close to the base of the jet, i.e. within $\sim 100\,\rin$. The optical depths also strongly depend on the observation angle of the accretion disk. These features are generally consistent with the early calculations for the accretion disks with
other parameters (see e.g. \citealp{car92, bed93}).
As a result of these strong absorption effects, the $\gamma$-ray spectra escaping to the observer should form in the process of the IC $e^\pm$ pair cascade developing above the disk surface. The $\gamma$-ray spectra emerging  from such a cascade should strongly depend on the location of the observer with respect to the axis of the disk, and the injection distance of primary electrons from the base of the jet as already considered in the case of the accretion disks in active galactic nuclei \citep{sb10a, sb10b}.

\section{IC $\electron^\pm$ pair cascade spectra}

In this section we investigate the IC $e^\pm$ pair cascade spectra as a function of the parameters describing the model. 
\kom{We perform a calculation of the pair cascade using a Monte Carlo code described in \citep{sb10b}. 
We expanded the original code to track also the spectra of the synchrotron radiation produced by electrons in the jet.
The jet velocity is assumed to be constant in its part, where the injection of electrons occurs.
The traverse size of the acceleration region is assumed to be much smaller then its height above the disk. 
However we also take into account a possibility that the blob is traveling along the jet resulting in an elongated acceleration region (see Section \ref{sec:whole}). }

\begin{figure}
\includegraphics[width=0.49\textwidth]{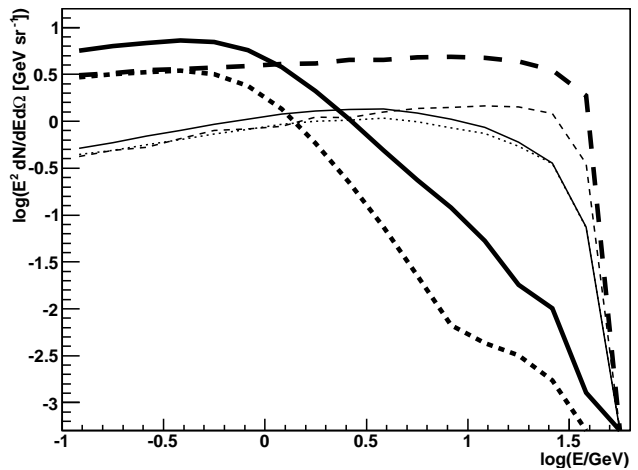}
\caption{
The $\gamma$-ray spectral energy distribution (SED) produced in the IC $e^\pm$ pair cascade (the solid curves) for two ranges of the observation angles  $0-40^\circ$ (the thin curves), and $60-75^\circ$ (thick).
The SEDs of the first generation of $\gamma$-rays produced by the primary electrons without any absorption taken into account are shown with the dashed curves. 
The SEDs of the $\gamma$-rays escaping from the binary system in the case of only simple exponential absorption law are shown with the dotted curves. 
It is assumed that the electrons are injected in a stationary blob at the height $H=100\,\rin$ above the disk. They have a differential power law spectrum with the index of $-2$ between 0.1~GeV and the maximum energy obtained from the comparison of energy gains and losses (Sect.~2.1). 
The magnetic field in the jet is described by the equipartition coefficient $\eta=10^{-6}$ and the decay law along the jet given by Eq.~1 (see text for details). Electron spectra are normalized to 1 erg.}
\label{fig_prod}
\end{figure}

At first we show what is the effect of the IC $e^\pm$ pair cascade process on the emerging $\gamma$-ray spectra in comparison to the first generation (primary) $\gamma$-rays produced in the cascade and also to the case of the simple exponential absorption law of these primary $\gamma$-rays (see Fig.~\ref{fig_prod}). 
The full cascade $\gamma$-ray spectra show a clear excess at energies below $\sim 1$ GeV (i.e. the threshold for the absorption process) in respect to the primary $\gamma$-ray spectra. 
This excess would have been neglected in the $\gamma$-ray spectra which are modified only by the simple exponential absorption law.
Moreover the simple exponential absorption cut-off in the $\gamma$-ray spectrum above $\sim 1$ GeV is partially filled up by the secondary cascade $\gamma$-rays produced by electrons with larger energies
(compare the thick dotted and the solid curves in Fig.~5). 
The full cascade spectrum steepens sharply in comparison to the primary $\gamma$-ray spectrum (which corresponds to the case of an optically thin, complete cooling IC $\gamma$-ray production). 
The primary $\gamma$-ray spectrum has a differential spectral index $\sim 2$, while the one of the full cascade spectrum above the break is closer to $3-4$. 
The effects mentioned above are especially visible in the case of the large observation angles, for which the cascading effects are the strongest due to the favourable interaction angles between the $\gamma$-rays and photons coming from the accretion disk.

\subsection{Spectra for fixed distances from the disk}\label{sec:fixed}

\begin{figure*}
\centering
%% scale=0.54 (normal)
%% \includegraphics[scale=0.54, trim=0  18 0 0, clip]{plots/synchkaskada_syn_eta1e-3_h1to1_g1.155_1.eps}
%% \includegraphics[scale=0.54, trim=31 18 0 0, clip]{plots/synchkaskada_syn_eta1e-3_h10to10_g1.155_1.eps}
%% \includegraphics[scale=0.54, trim=31 18 0 0, clip]{plots/synchkaskada_syn_eta1e-3_h100to100_g1.155_1.eps}
%% \includegraphics[scale=0.54, trim=0 0 0 0  , clip]{plots/synchkaskada_syn_eta1e-6_h1to1_g1.155_1.eps}
%% \includegraphics[scale=0.54, trim=31 0 0 0 , clip]{plots/synchkaskada_syn_eta1e-6_h10to10_g1.155_1.eps}
%% \includegraphics[scale=0.54, trim=31 0 0 0 , clip]{plots/synchkaskada_syn_eta1e-6_h100to100_g1.155_1.eps}
\includegraphics[width=0.349\textwidth, trim=0  18 0 0, clip]{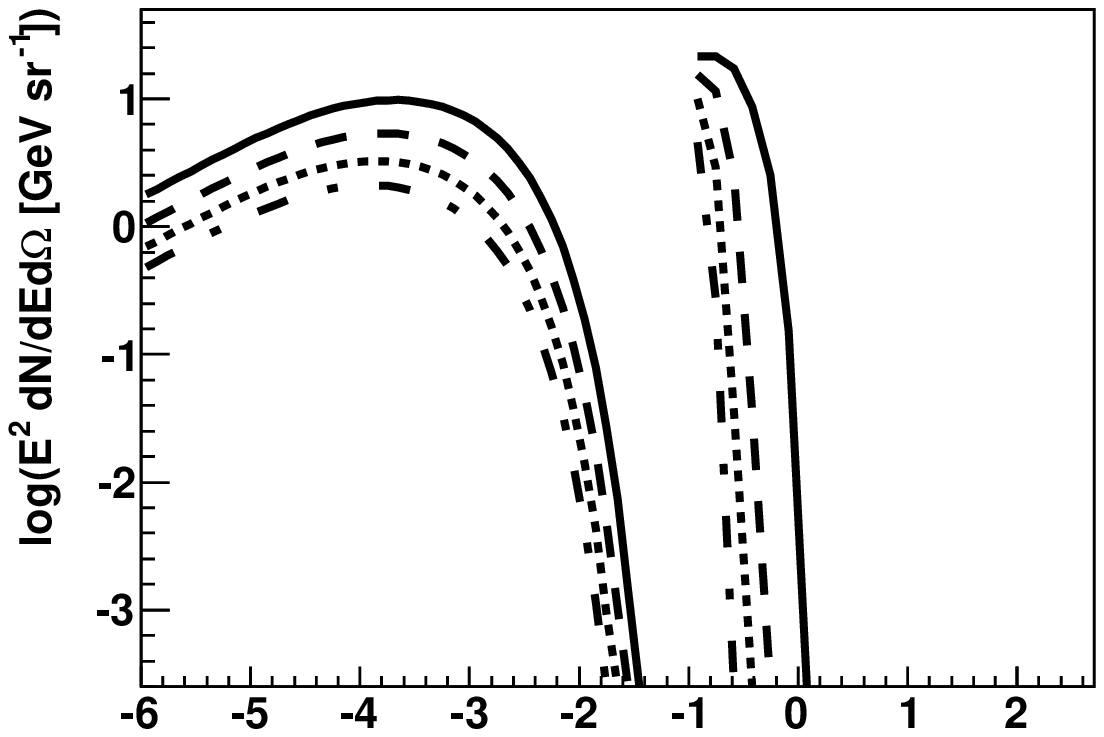}
\includegraphics[width=0.315\textwidth, trim=31 18 0 0, clip]{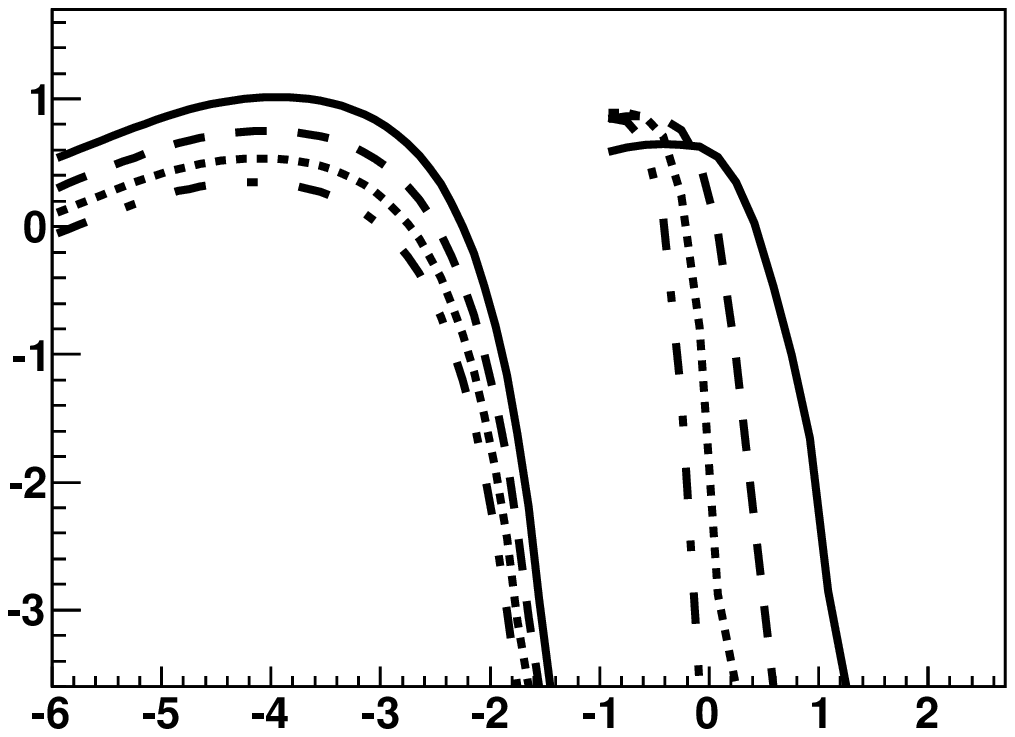}
\includegraphics[width=0.315\textwidth, trim=31 18 0 0, clip]{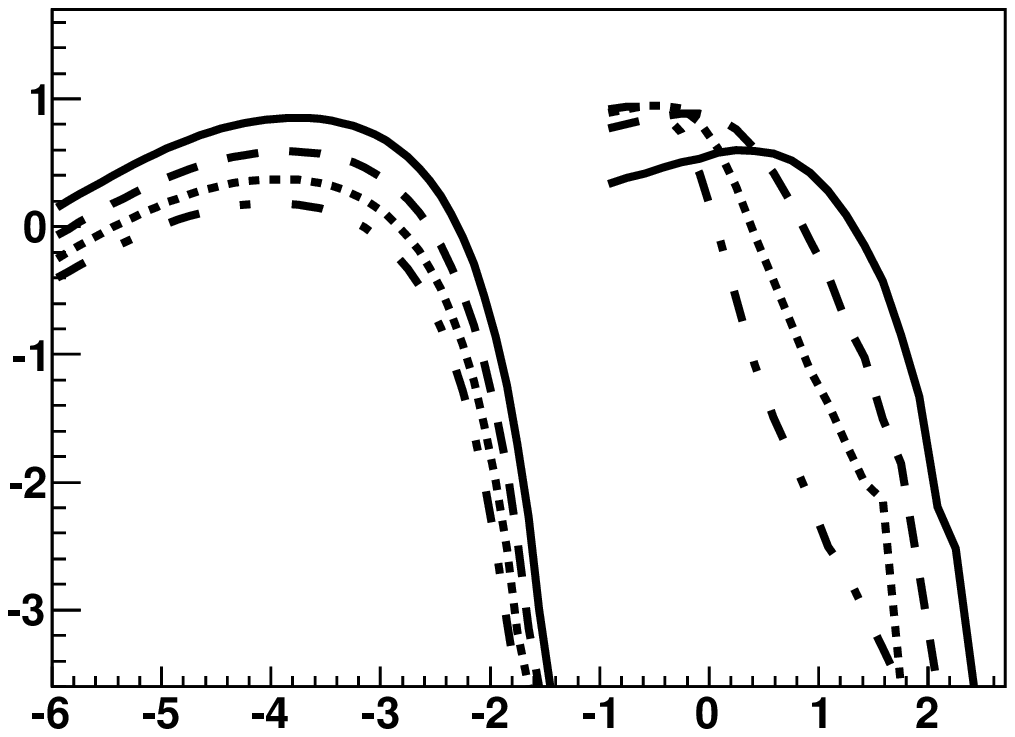}
\includegraphics[width=0.349\textwidth, trim=0 0 0 0  , clip]{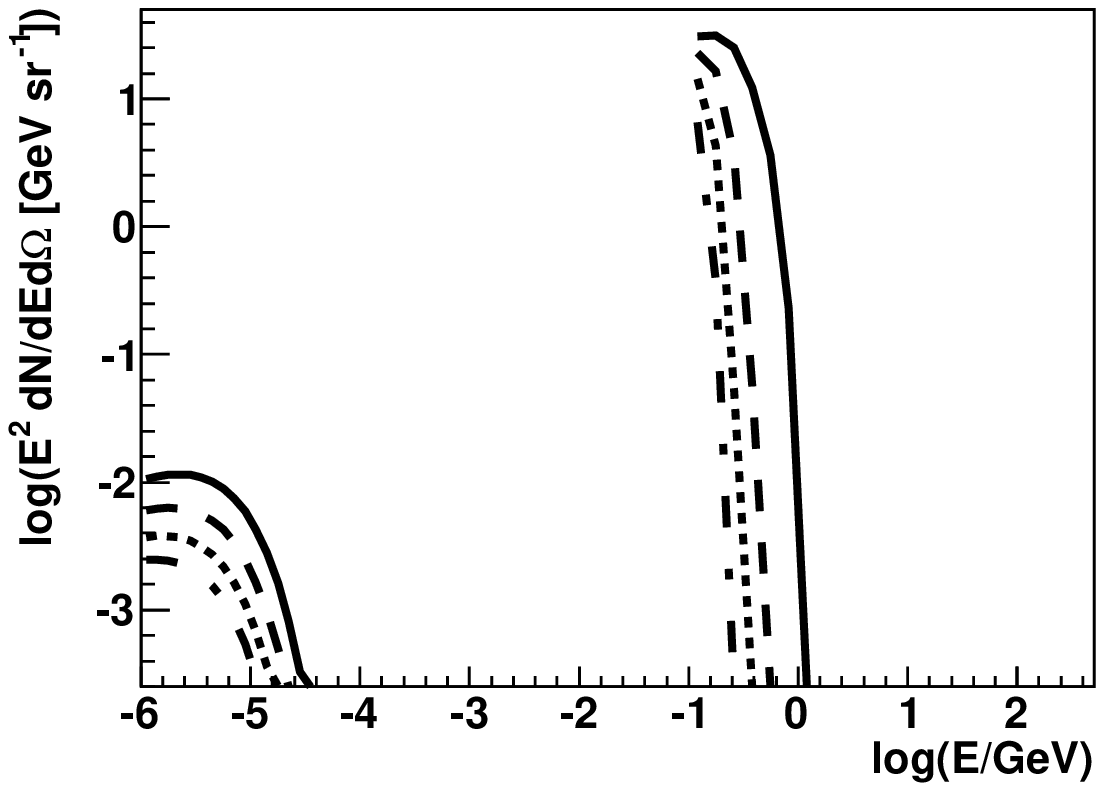}
\includegraphics[width=0.315\textwidth, trim=31 0 0 0 , clip]{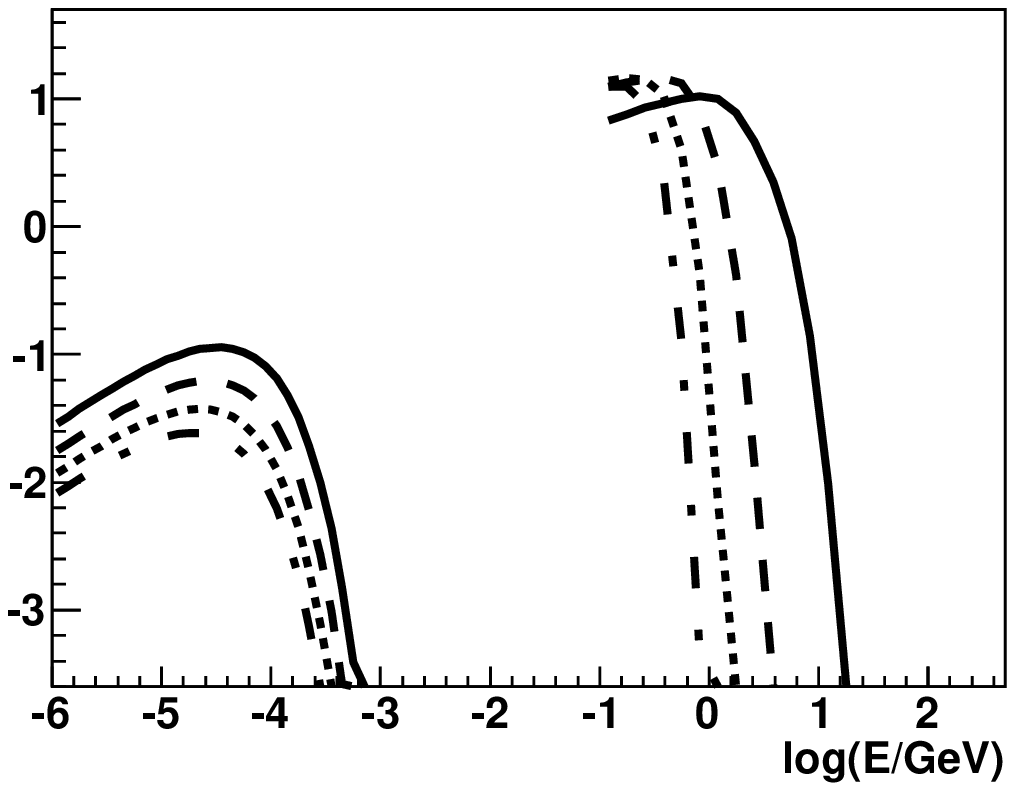}
\includegraphics[width=0.315\textwidth, trim=31 0 0 0 , clip]{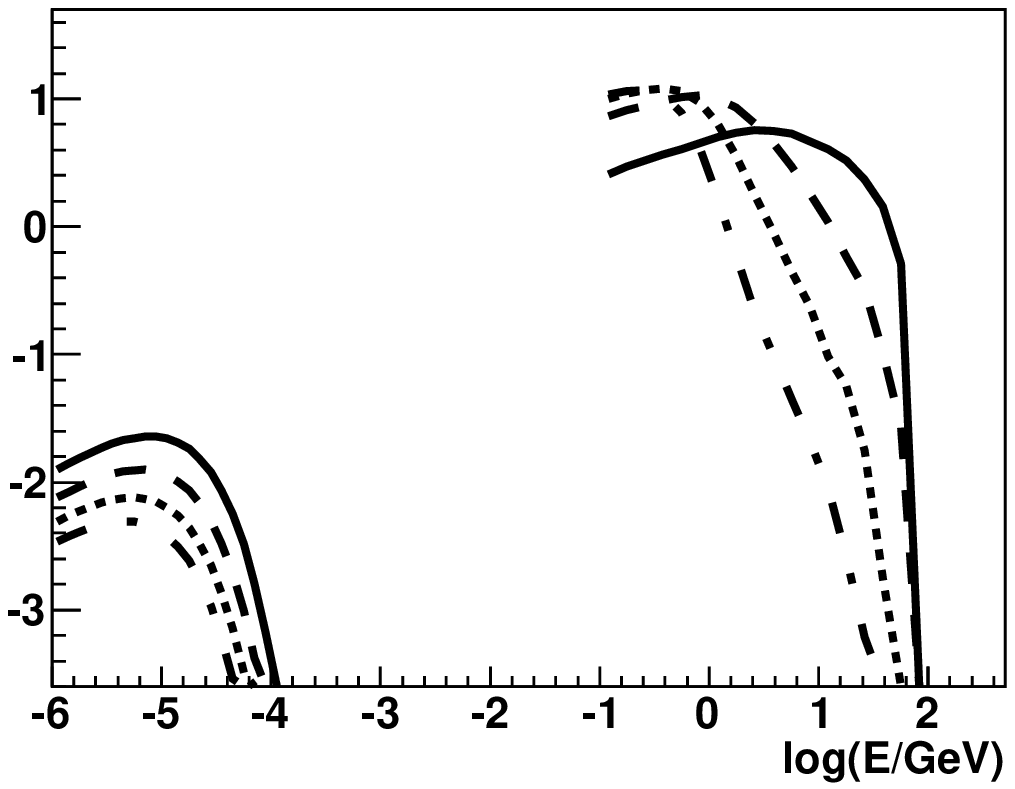}
\caption{$\gamma$-ray SED, produced in the IC $e^\pm$ pair cascade \kom{(rightmost component, only calculated $>0.1\,$GeV)}, observed at different range of angles measured in respect to the jet axis:
$0-40^\circ$ (the solid curves),
$40-60^\circ$ (dashed),
$60-75^\circ$ (dotted),
and $75-90^\circ$ (dot-dashed).
\kom{Leftmost component of the SEDs is produced by simultaneous synchrotron radiation of the primary electrons.}
The primary electrons are injected with the differential power law spectrum with an index $-2$ between 0.1~GeV and the maximum energy determined by balancing energy gains and losses (see Sect.~2.1). 
They are injected from a blob moving with the velocity $v_{\rm b} = 0.5$c at its fixed distance from the base of the jet: $H = 1\,\rin$ (the left panels), $10\,\rin$ (the middle panels), $100\,\rin$ (the right panels). 
The calculations have been performed for the acceleration coefficient equal to $\xi$=0.01 and the equipartition parameter describing the magnetic field strength in the blob equal to $\eta=10^{-3}$ (the top panels) and $10^{-6}$ (the bottom panels). 
}
\label{fig_fixed}
\end{figure*}

We have calculated the $\gamma$-ray spectra produced in the cascade process by electrons injected in the blob moving with the specific velocity,
$v_{\rm b} = 0.5$c, and located at a fixed distance from the accretion disk (see Fig.~\ref{fig_fixed}). 
In these calculations electrons are injected into the blob at a given location in the jet but after the injection process the blob changes its distance from the disk. 
\kom{As the calculations of a cascade in an anisotropic radiation field are very time and memory consuming we calculate the high energy component (from the IC scattering) only down to $0.1\,$GeV, which is the energy range accessible by the Fermi-LAT instrument. 
Presently there is no sensitive enough experiment to investigate the MeV range. 
Moreover in order to accurately predict the MeV emission (the joining point of the synchrotron and IC component), one would have to do additional assumption on the acceleration mechanism (e.g. what is the minimal energy, from which electrons are accelerated).
}

Two scenarios for the magnetic field distribution in the jet are considered. In the first one, model~I (defined by the equipartition parameter equal to $\eta=10^{-3}$), the energy losses of electrons on the synchrotron and IC processes are comparable.  In the second one, model~II  ($\eta = 10^{-6}$), the energy losses on the IC process clearly dominate over their synchrotron losses. 
In the model~I, the synchrotron spectra produced by the primary electrons in the blob almost do not depend on the distance from the base of the disk. 
This feature is easy to understand keeping in mind that maximum energies of electrons accelerated in specific parts of the jet are determined also by the balance between the energy gains and the synchrotron energy losses. 
As a result the maximum energies of synchrotron photons do not change with different strength of the magnetic field. They depend only on the acceleration parameter. 
In the model~II, the synchrotron spectra show complicated dependence on the distance from the disk since the maximum energies of accelerated electrons are determined in this case by their IC energy losses (see Fig.~3). The $\gamma$-ray spectra from the IC $e^\pm$ pair cascade process in both models clearly extend to larger energies for larger injection distances from the base of the disk.
In general, the $\gamma$-ray spectra cut-off at lower energies for larger observation angles due to stronger absorption effects for $\gamma$-rays 
propagating at larger angles to the disk axis (see Fig.~4). 
Note the characteristic shapes of the cascade $\gamma$-ray spectra escaping from the disk radiation. 
They show a clear anticorrelation in energy ranges below and above a few GeV with the observation angle, i.e. fluxes above/below  a few GeV are relatively large/small for small angles but for large angles the situation is the opposite.
As expected, the multiwavelength spectra in such a model for a weakly magnetized jet (i.e. the magnetic field strength at the bottom of the jet $B_{\rm in}\sim 10^4$ G) show strong dominance of emission at $\gamma$-ray energies.
Such sources should be clearly detected in $\gamma$-ray energy range showing relatively weak synchrotron emission in the range from radio to X-rays. 
In contrary, in strongly magnetized jets ($B_{\rm in}\sim 3\times 10^5$ G), the power emitted in the synchrotron and in the IC bumps are comparable. 
We show quantitatively that microquasars can show completely different spectral features depending on the magnetization parameter of the jet.

\subsection{Spectra from the whole jet}\label{sec:whole}

\begin{figure*}
\centering
\includegraphics[width=0.263\textwidth, trim=0  18 0 0, clip]{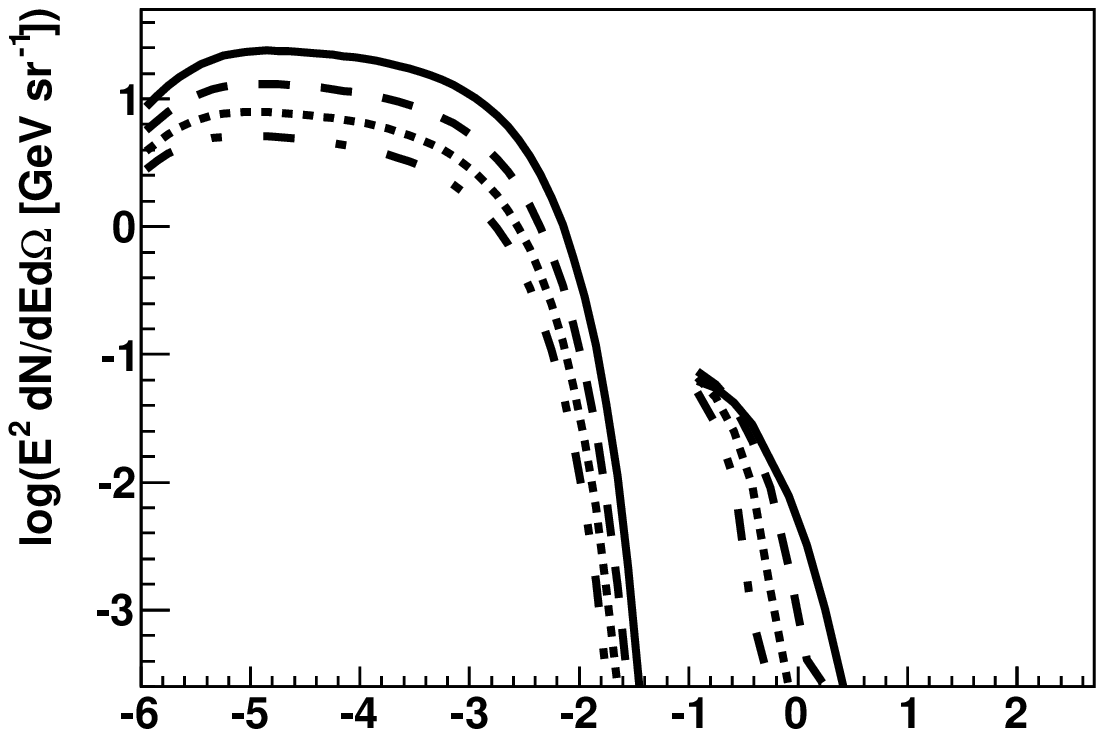}
\includegraphics[width=0.237\textwidth, trim=31 18 0 0, clip]{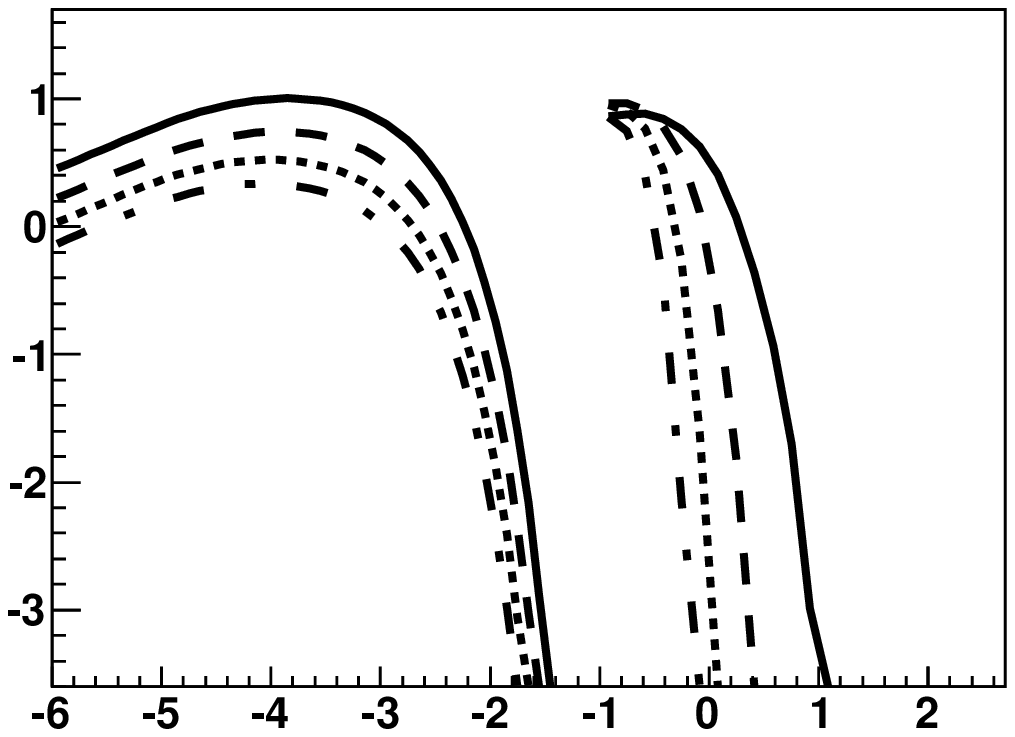}
\includegraphics[width=0.237\textwidth, trim=31 18 0 0, clip]{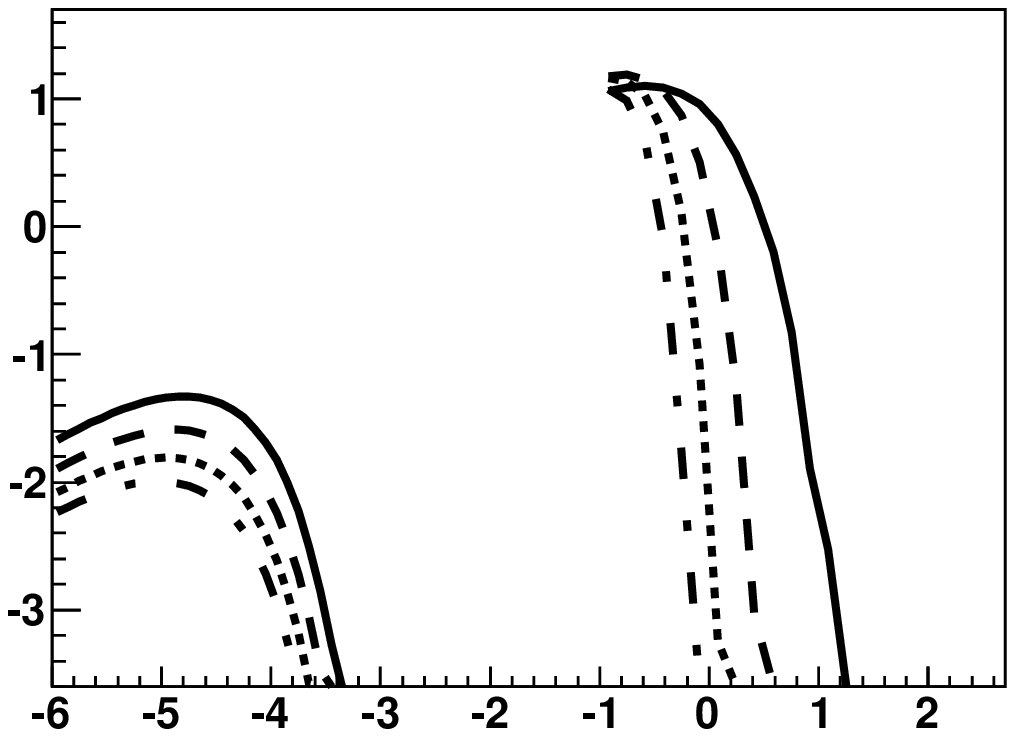} 
\includegraphics[width=0.237\textwidth, trim=31 18 0 0, clip]{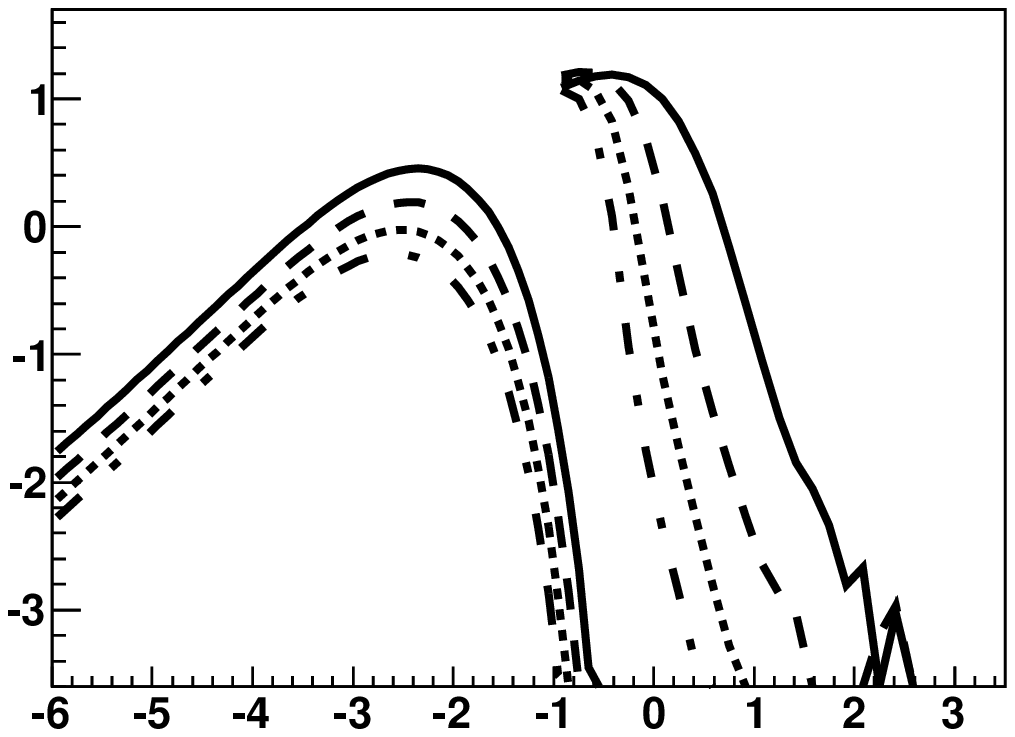}\\
\includegraphics[width=0.263\textwidth, trim=0   0 0 0, clip]{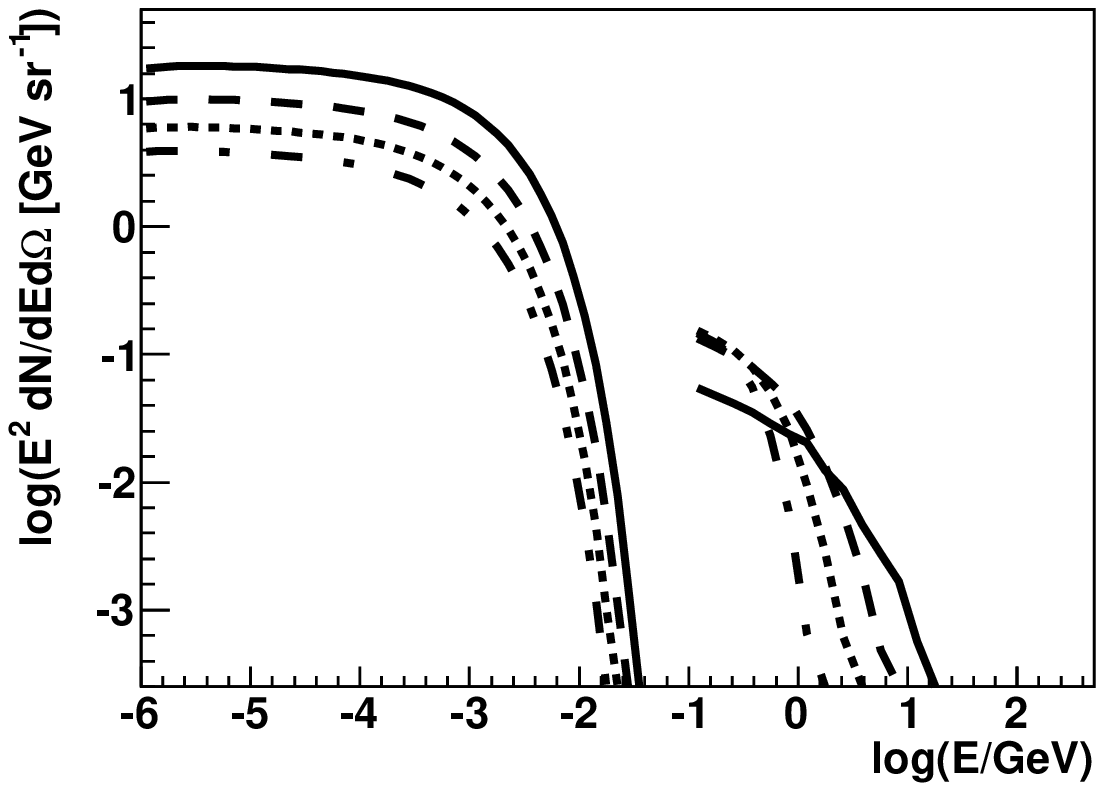}
\includegraphics[width=0.237\textwidth, trim=31  0 0 0, clip]{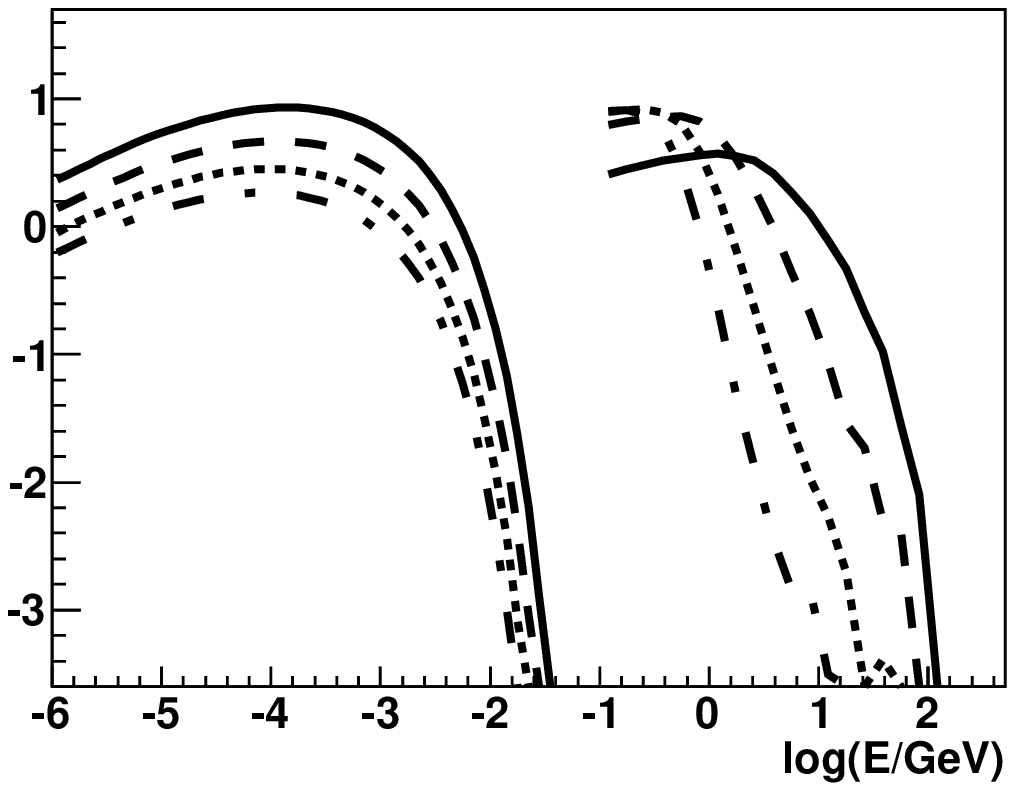}
\includegraphics[width=0.237\textwidth, trim=31  0 0 0, clip]{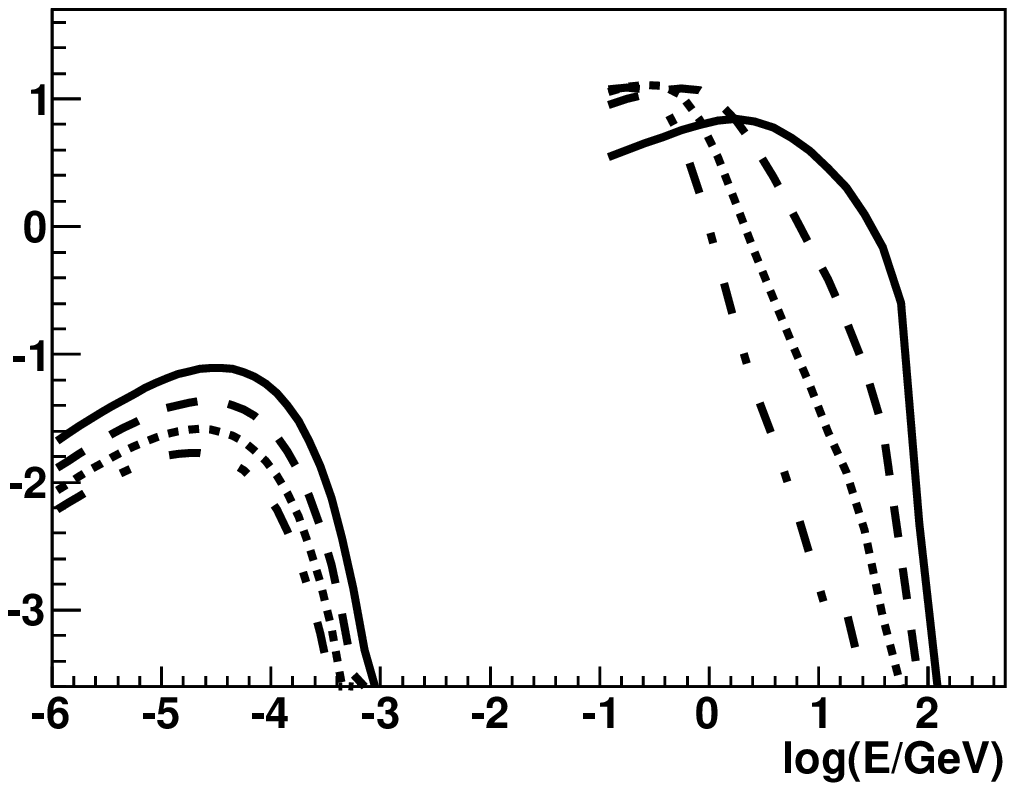}
\includegraphics[width=0.237\textwidth, trim=31  0 0 0, clip]{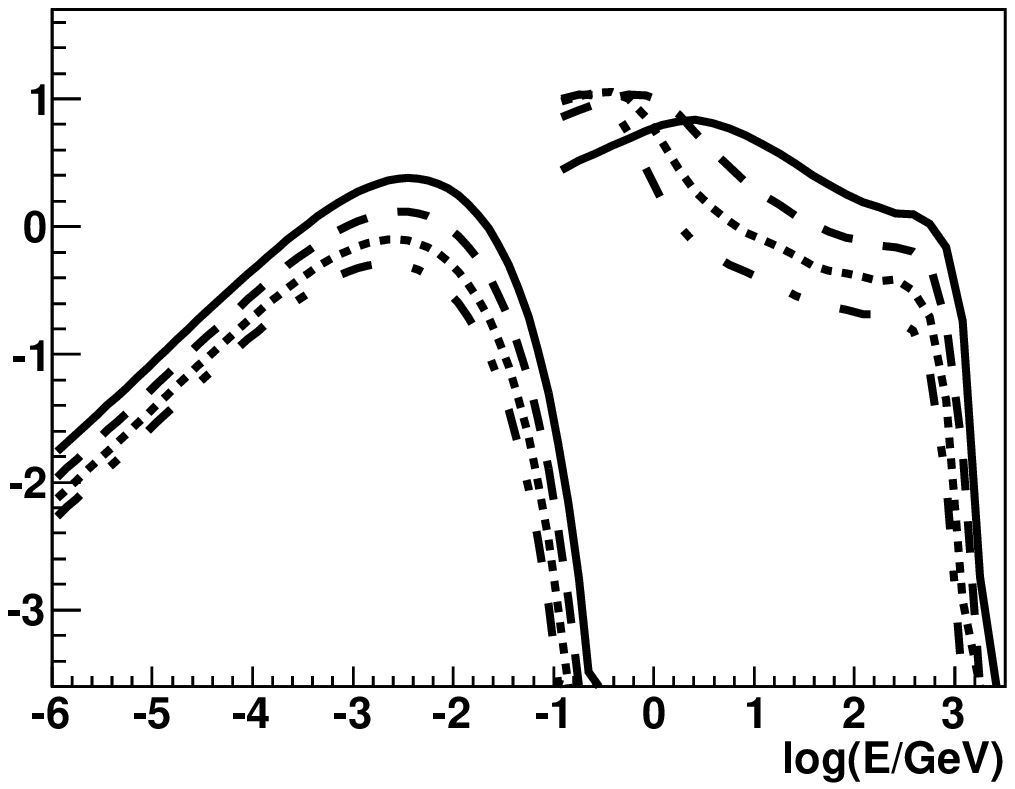}
\caption{
As in Fig.~\ref{fig_fixed}, but for the continuous injection of electrons along the jet through the range of distances from the disk: 
$H = 1-10\,\rin$ (the top panels), $10-100\,\rin$ (the bottom panels).
The equipartition parameter and the acceleration efficiency are fixed on: $\eta=1, \xi=0.01$ (the far left panels) $\eta=10^{-3}, \xi=0.01$ (the middle left panels), $\eta=10^{-6}, \xi=0.01$ (the middle right panels) and $\eta=10^{-6}, \xi=0.1$ (the far right panels). 
The velocity of the blob is assumed to be $0.5c$.
\kom{Rightmost component of the SED is produced by the IC scattering in the $e^\pm$ pair cascade (only calculated $>0.1\,$GeV), while the leftmost component is the simultaneous synchrotron radiation of the primary electrons.}
}
\label{fig_range}
\end{figure*}

We also calculate the $\gamma$-ray spectra by assuming that the injection of electrons occurs not only at a specific distance from the base of the jet but rather extends along the considerable part of the jet. 
This corresponds to the situation in which the acceleration process of the electrons operates through a certain period of time during which the blob passes specific part of the jet (see Fig.~\ref{fig_range}).
It is assumed that during this period the injection rate of electrons is constant but the maximum energies of freshly injected electrons are determined by the balance between the acceleration efficiency and the energy losses at the specific distance from the disk. 
We also follow the cooling of earlier injected electrons during the propagation of the blob within the jet. As an example, we investigate how the synchrotron and the cascade $\gamma$-ray spectra depend on the basic parameters of the model for two ranges of injection distances from the disk ($H = 1-10\,\rin$ and $10-100\,\rin$). 
As it can be expected from Sect.~\ref{sec:fixed}, depending on the magnetization parameter of the jet ($\eta$) and the acceleration efficiency of electrons ($\xi$) we predict completely different shapes for the multiwavelength spectra at high energies.
Either the synchrotron spectra dominate in the energy range from soft X-rays up to soft $\gamma$-ray (without significant emission at $\sim 100$ MeV) or the IC $\gamma$-ray spectra dominate with relatively weak fluxes at the synchrotron part of the spectrum which peak usually at hard X-rays. 
Note that depending on the discussed scenario, the spectral index of the differential synchrotron spectrum changes from $\sim -2$ (for the synchrotron cooling dominated regime) to $\sim -1.3$ (for the IC cooling dominated regime) (see Fig.~\ref{fig_range}).

The increase of the acceleration parameter to the maximum possible values expected in our model ($\xi\sim 0.1$) allows the acceleration of electrons to energies with which they are able to produce the IC $\gamma$-ray spectra extending up to $\sim 1$ TeV energies (i.e. in the range well covered by the presently operating Cherenkov telescopes). 
We suggest that some of the unidentified galactic TeV $\gamma$-ray sources (without obvious counterpart in the radio and X-rays) might be in fact  low mass microquasars efficiently accelerating electrons. 

\begin{figure}
%\centering
\includegraphics[scale=0.385, trim=0  18 0 0, clip]{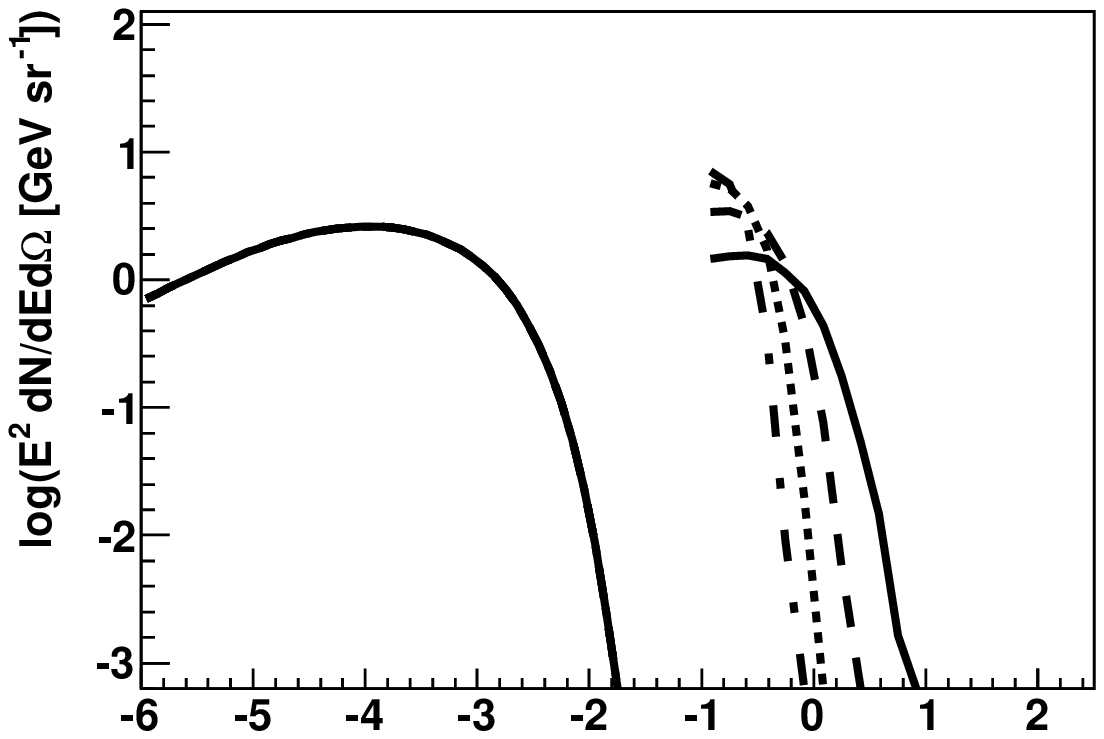}
\includegraphics[scale=0.385, trim=31 18 0 0, clip]{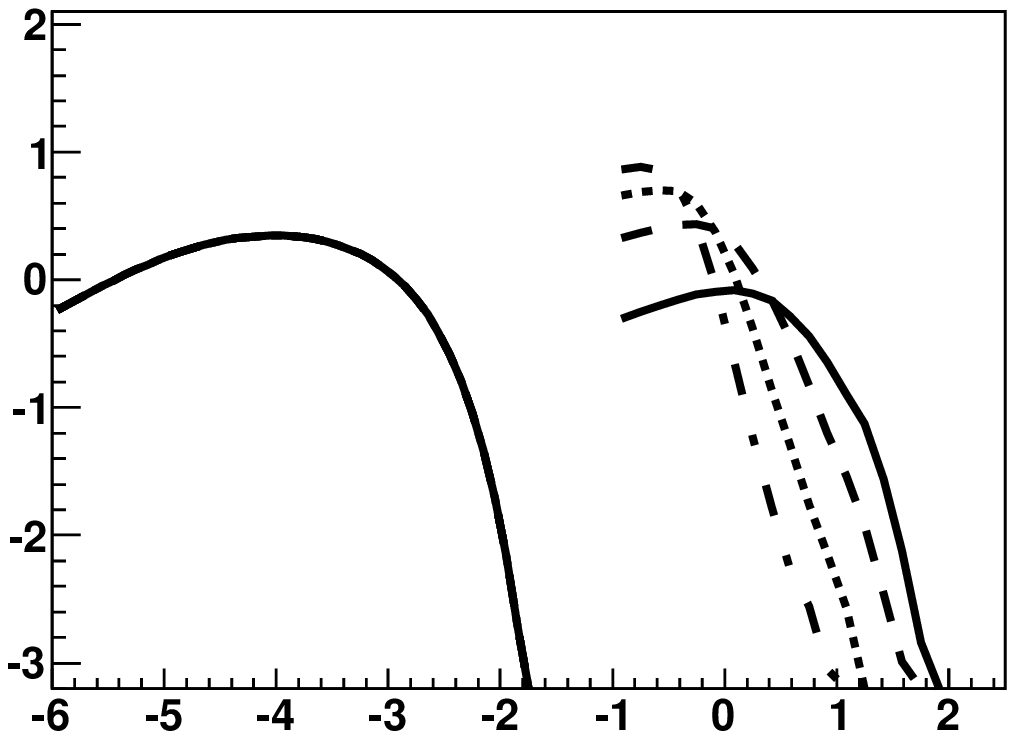}\\
\includegraphics[scale=0.385, trim=0  18 0 0, clip]{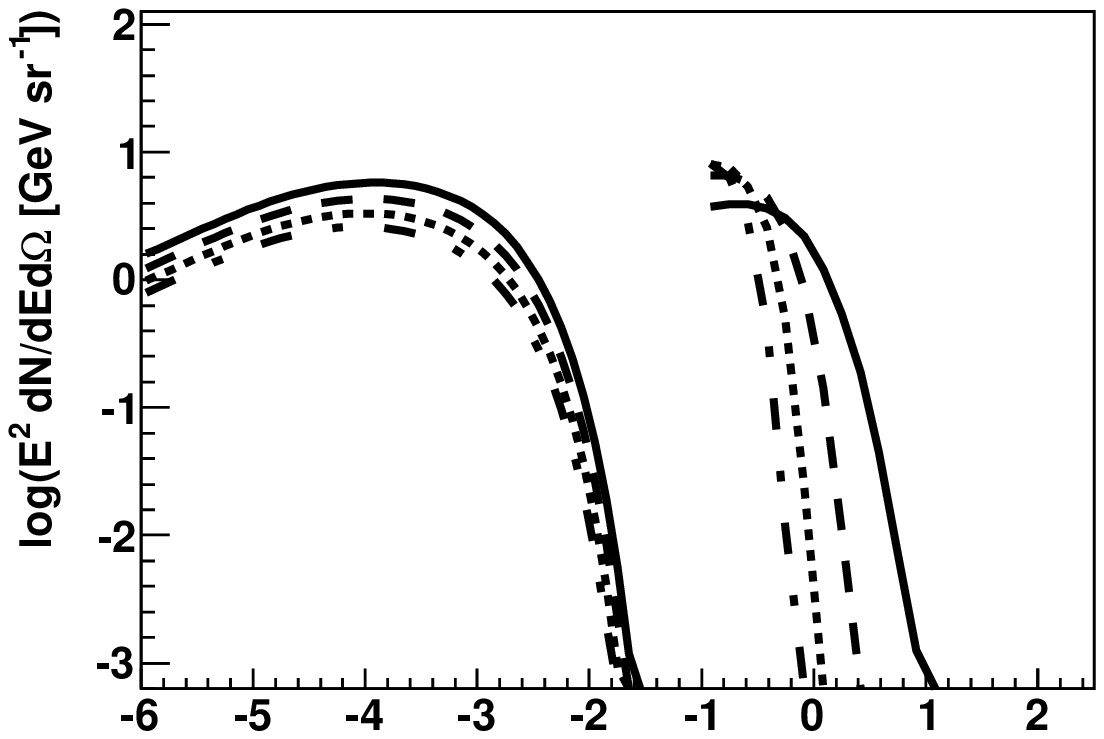}
\includegraphics[scale=0.385, trim=31 18 0 0, clip]{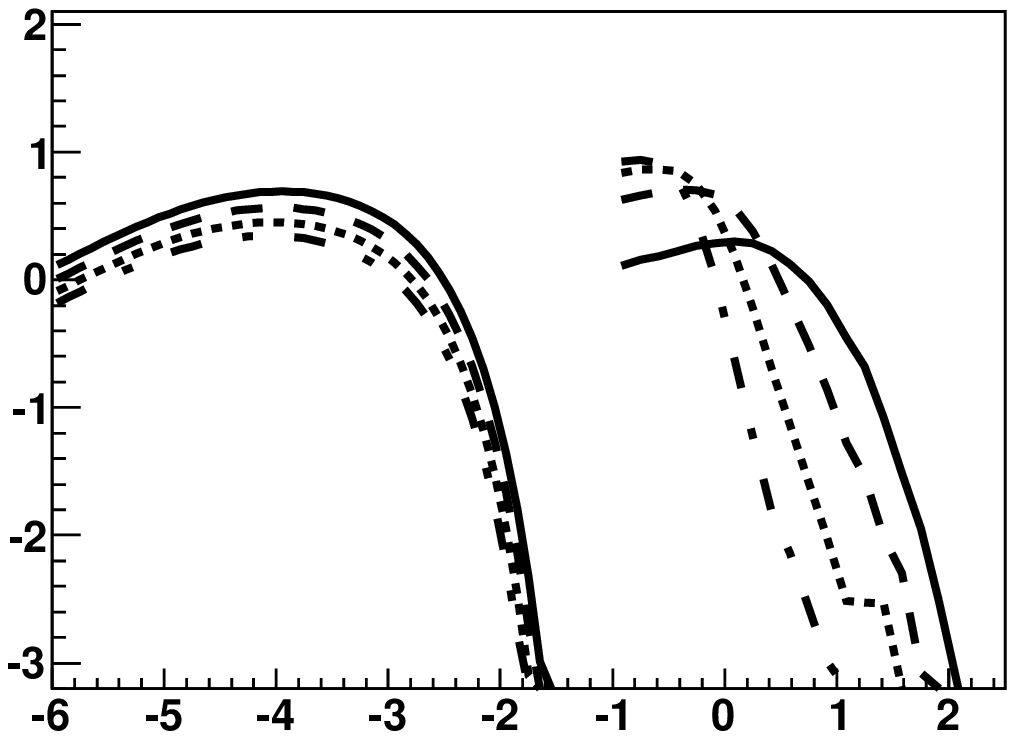}\\
\includegraphics[scale=0.385, trim=0   0 0 0, clip]{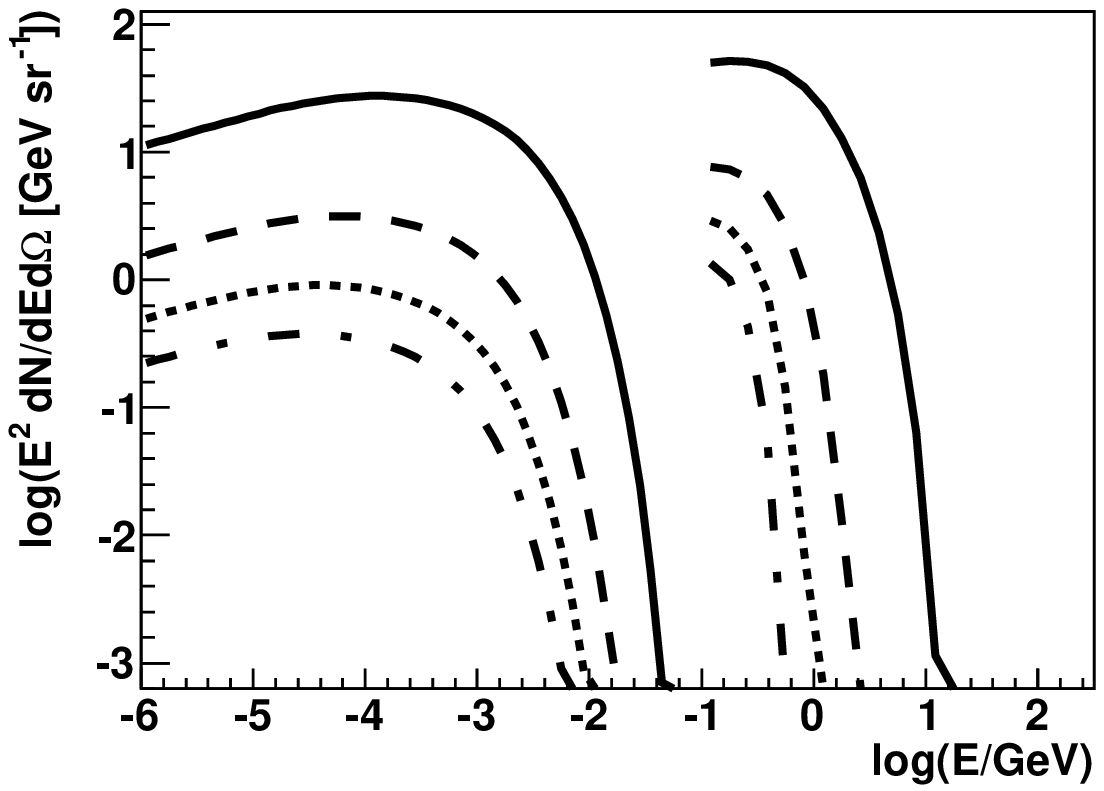}
\includegraphics[scale=0.385, trim=31  0 0 0, clip]{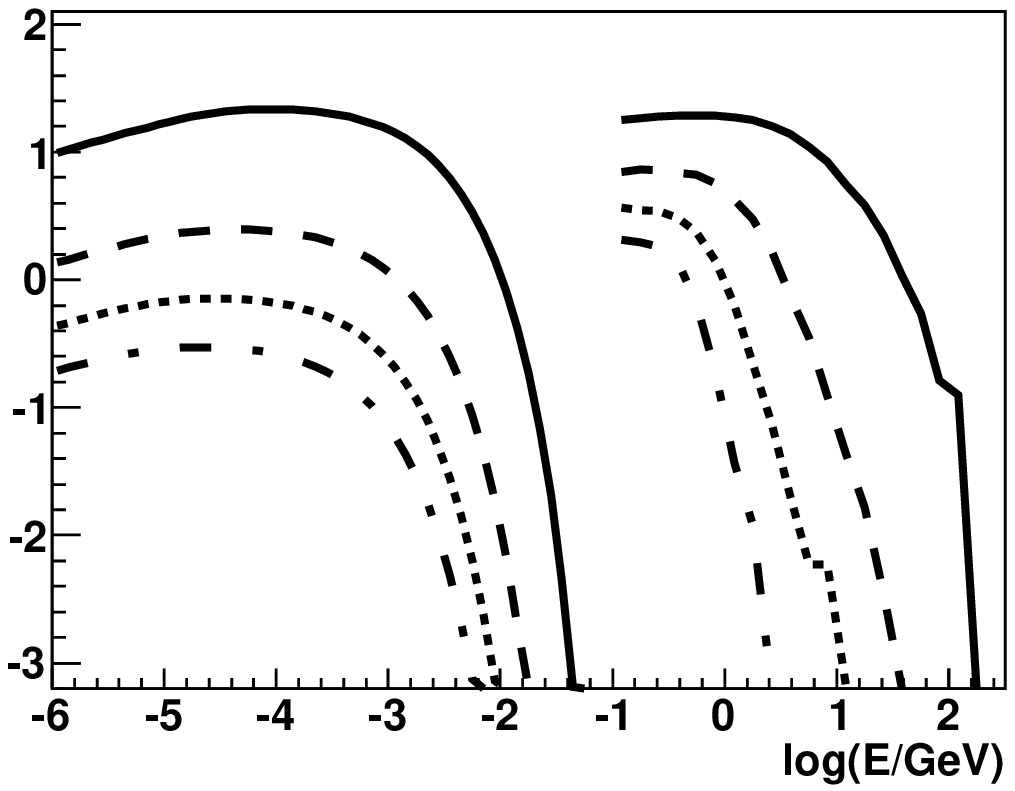}
\caption{As in Fig.~\ref{fig_range} but for different velocities of the blob in the jet: $v_b=0$ (emitting region is at rest, the top panels), $v_b=0.3\,c$ (the middle panels) and $v_b=0.9\,c$ (the bottom panels).
Injection occurs at the height $H = 1-10\,\rin$ (the left panels), $10-100\,\rin$ (the right panels). 
The equipartition parameter and the acceleration efficiency is fixed on: $\eta=10^{-3}$, $\xi=0.01$ respectively.
\kom{Rightmost component of the SED is produced by the IC scattering in the $e^\pm$ pair cascade (only calculated $>0.1\,$GeV), while the leftmost component is the simultaneous synchrotron radiation of the primary electrons.}
}
\label{fig_speed}
\end{figure}

The IC cascade $\gamma$-ray spectra discussed above have been performed for the fixed velocity of the blob along the jet. 
In Fig.~\ref{fig_speed} we show how these $\gamma$-ray spectra depend on the velocity of the blob for the range of values observed in microquasars, i.e. $0.3c-0.9c$. 
As expected, the relativistic Doppler effect is important for the jets with large velocities.
Then, the $\gamma$-ray spectra emitted at small observation angles to the jet axis clearly dominate over those emitted at large angles even in the sub-GeV energies. Therefore, in principle the shape of the $\gamma$-ray spectrum observed through broad energy range should give some independent constraints on the velocity of the $\gamma$-ray production region in microquasars. Note that for large velocity jets, the spectra also extend to larger energies, which increases the chances for a detection of such sources by the Cherenkov telescopes.

\subsection{Spectra for different accretion disks}

\begin{figure*}
\includegraphics[width=0.343\textwidth, trim= 0 0 5 0, clip]{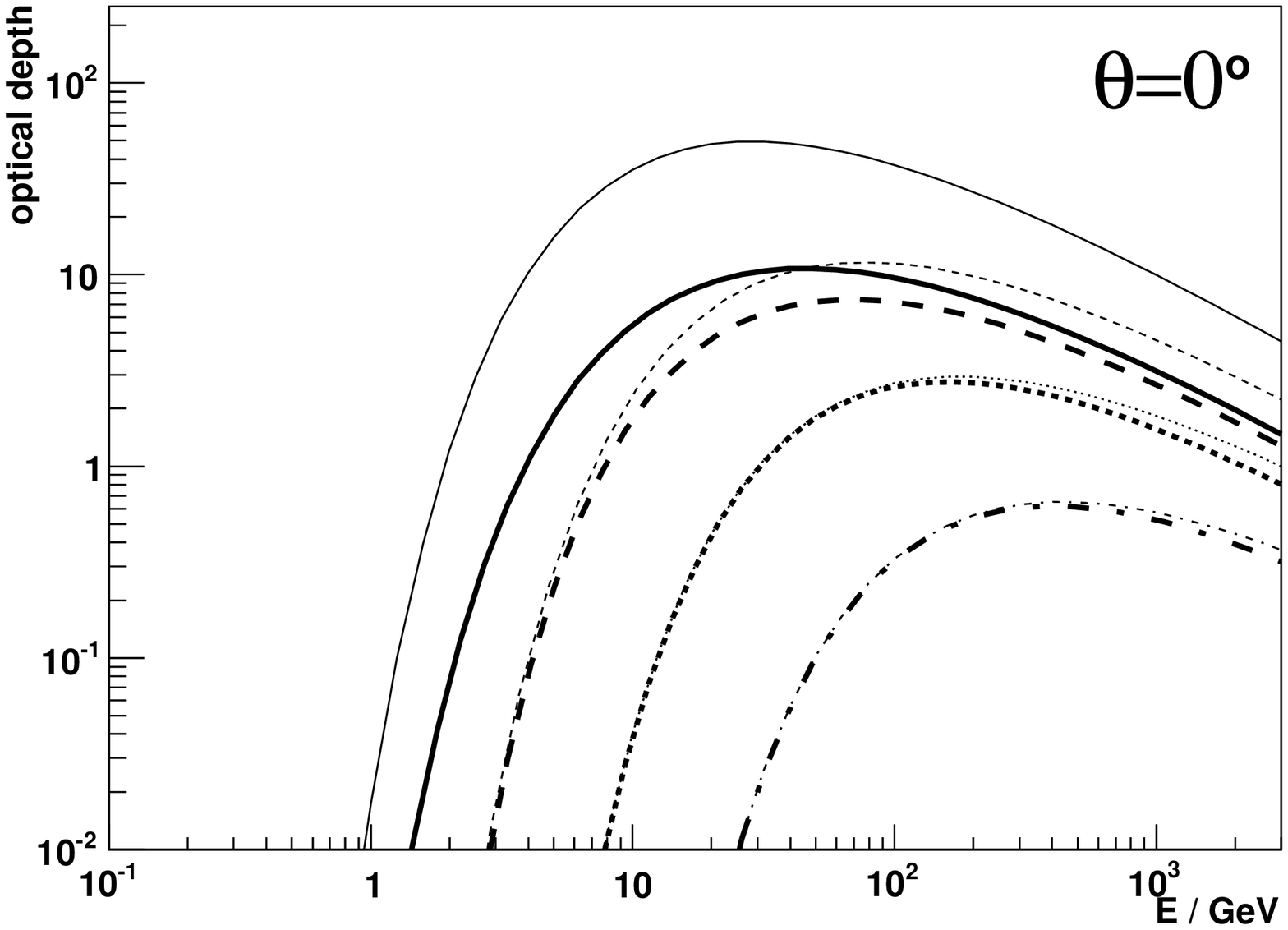}
\includegraphics[width=0.318\textwidth, trim=40 0 5 0, clip]{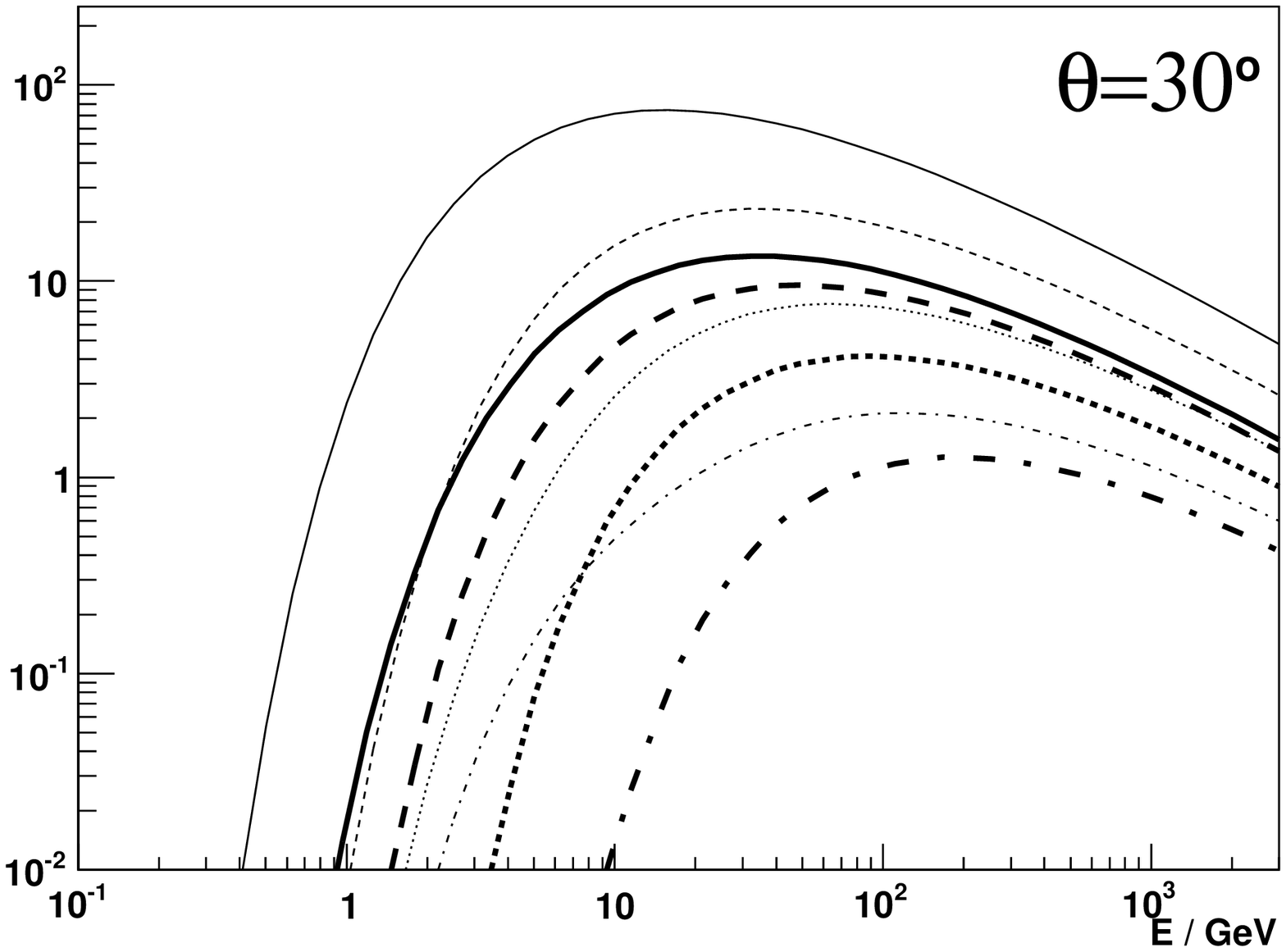}
\includegraphics[width=0.318\textwidth, trim=40 0 5 0, clip]{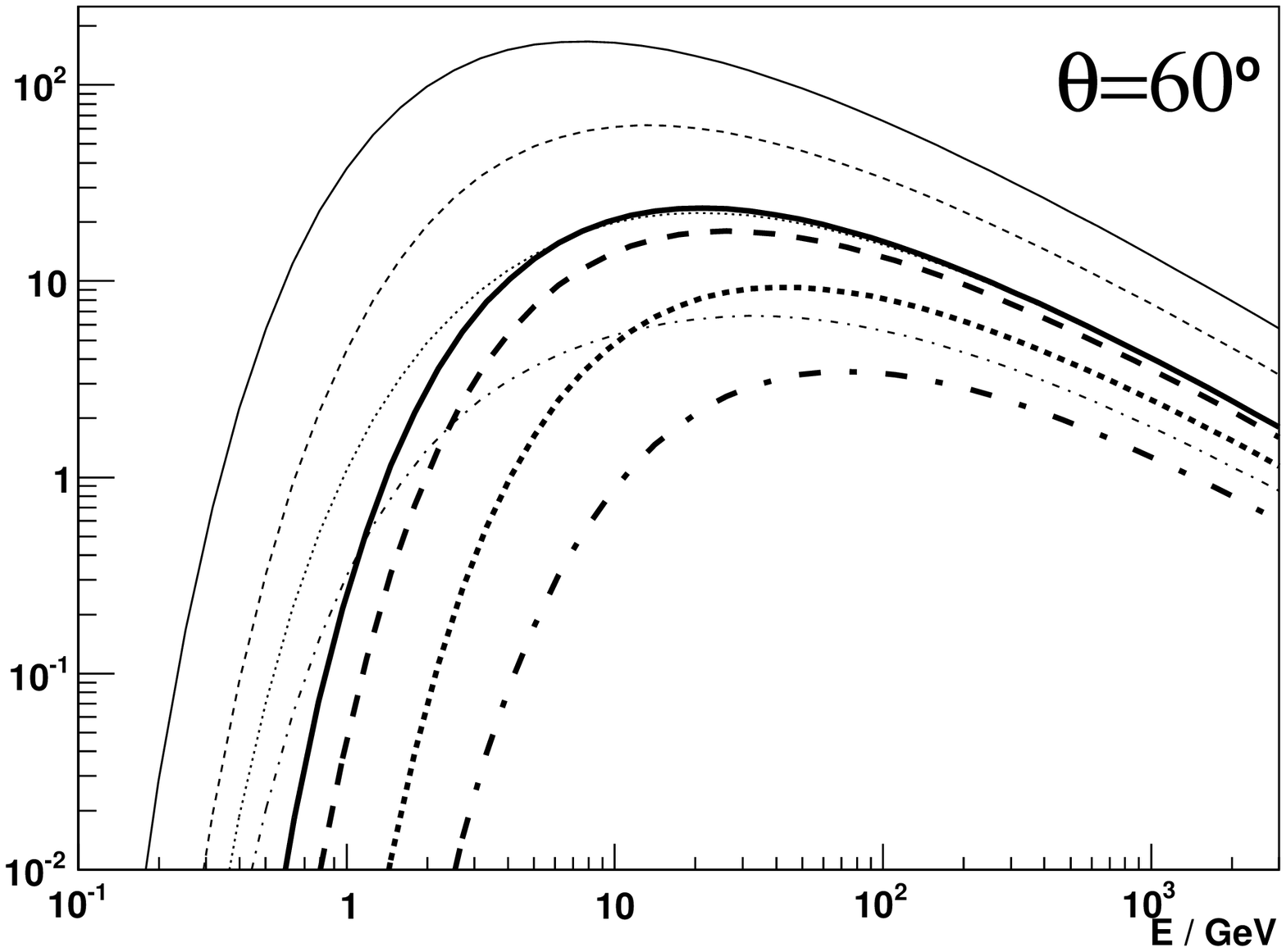}
\caption{
The optical depths for $\gamma$-rays in the radiation field of the accretion disk assuming that the inner part of the disk becomes radiatively inefficient (the radiation from this part of the disk is neglected).
The thin curves marks the results for the standard Shakura \& Sunyaev accretion disk with temperature $\tin=5\cdot10^{6}$~K at the inner radius of the disk $\rin=10^{7}$~cm.
The thick curves are the optical depths for the disk in which the emission is generated only above $10\,\rin$.
Different curve styles correspond to different heights above the disk: 
$H=3\,\rin$ (solid), $10\,\rin$ (dashed), $30\,\rin$ (dotted), and $100\,\rin$ (dot-dashed).
$\gamma$-ray photons are observed at the angles: $\theta=0^\circ$ (the left panel), $30^\circ$ (middle) and $60^\circ$ (right) with respect to the jet axis.}
\label{fig_taucut}
\end{figure*}

Up to now we considered only the Shakura-Sunyaev optically thick and geometrically thin accretion disks. In fact, the inner part of the accretion disk can become radiatively inefficient (the so-called advection dominated disks, e.g. \citealp{ny94}). 
In such case, the matter in the inner part of the accretion disk can become optically thin and the temperature of the plasma in the disk can rise by orders of magnitude. Such disk becomes also radiatively inefficient, i.e a significant part of the generated gravitational energy is swallowed by the black hole.
Then, the rare, hot plasma in the inner part of the disk produces radiation field, which is not dense enough neither for an efficient absorption of $\gamma$-rays, nor for $\gamma$-ray production via the IC scattering process. 
Therefore, we neglect this inner part of the disk when considering the  IC $e^\pm$ pair cascade processes. 
%For comparison, we have calculated the optical depths for $\gamma$-ray photons, for different injection angles, assuming that the disk inner radius is larger than previously considered value $\rin = 10^7$ cm (see Fig.~\ref{fig_taucut}). 
For comparison, we have calculated the optical depths for $\gamma$-ray photons, for different injection angles, assuming that the inner part of the accretion flow does not contribute to the absorption of the $\gamma$-rays (see Fig.~\ref{fig_taucut}). 
Note significant differences in the optical depths for $\gamma$-rays both for small and for large angles in respect to the disk axis.
The differences in the optical depths in respect to the previously discussed accretion disk are especially important at low energies due to the lack of the radiation from the inner, hotter part of the accretion disk.

\begin{figure}
\includegraphics[scale=0.39, trim=0  0 0 0, clip]{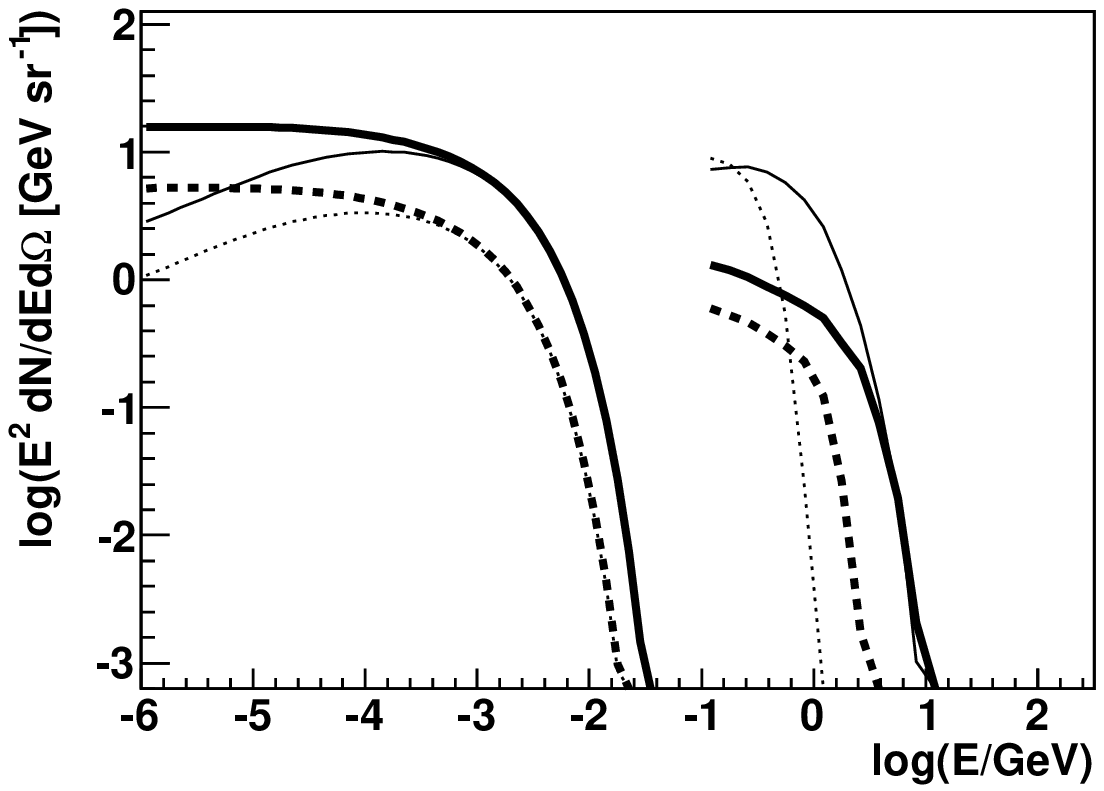}
\includegraphics[scale=0.39, trim=36  0 0 0, clip]{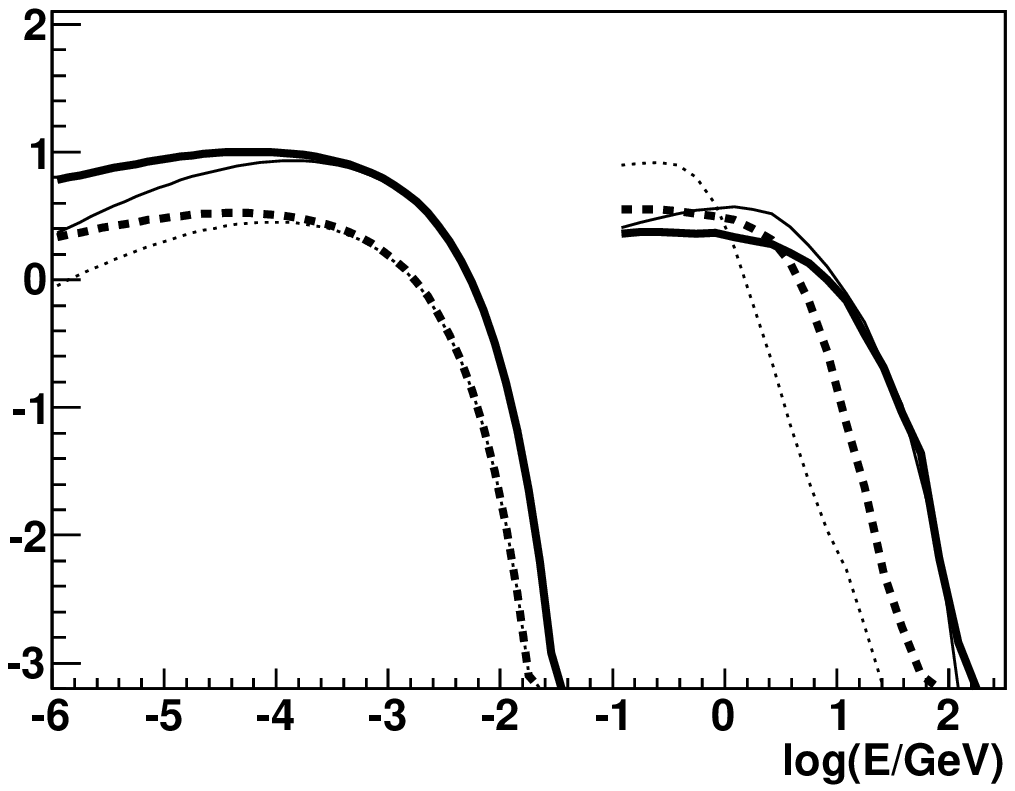}
\caption{
$\gamma$-ray SED produced in the IC $e^\pm$ pair cascade observed at different range of angles measured in respect to the jet axis:
$0-40^\circ$ (the solid curves),
and $60-75^\circ$ (dotted).
The spectra from the standard \citeauthor{ss73} disk (as the curves in Fig.~\ref{fig_range}) are shown by the thin curves, and from the truncated disk (without the inner part of the disk, below  $10\,\rin$) by the thick curves. 
The primary electrons are injected with the differential power law spectrum with an index $-2$ between 0.1~GeV and the maximum energy determined by balancing energy gains and losses (see Sect.~2.1) in the range of distances from the disk:  $H = 1-10\,\rin$ (the left panel), $10-100\,\rin$ (the right panel).
The blob is moving with the velocity $v_{\rm b} = 0.5c$.
The acceleration coefficient is equal to $\xi=0.01$ and the equipartition parameter is equal to $\eta=10^{-3}$.
\kom{Rightmost component of the SED is produced by the IC scattering in the $e^\pm$ pair cascade (only calculated $>0.1\,$GeV), while the leftmost component is the simultaneous synchrotron radiation of the primary electrons.}
} 
\label{fig_diskcut}
\end{figure}

In Fig.~\ref{fig_diskcut}, we compare the $\gamma$-ray spectra produced in the IC $e^\pm$ pair cascade, developing in the radiation field of such a truncated disk with the case of the whole \citeauthor{ss73} disk.
If the injection of the relativistic electrons occurs far away from the disk, the cascade $\gamma$-ray spectra observed for small viewing angles are very similar in both cases. 
The inner part of the accretion disk does not contribute significantly to the absorption process of $\gamma$-rays
due to the small angles between the directions of the $\gamma$-rays and the direction to the centre of the disk. 
On the other hand, the spectra produced at a large angles are extending to higher energies for radiatively inefficient disk due to the lack of the strong absorption in the hot radiation emitted from the central parts of the disk. 
In contrary, if electrons are accelerated close to the surface of the disk, then the lack of the emission from the inner part of the disk significantly reduce the level of the $\gamma$-ray spectra. 
Also the synchrotron spectra of the primary electrons are modified, especially for the injection close to the base of the jet, due to different relative efficiency of the two basic cooling processes considered in our model. Without the emission from the hot, inner part of the disk, the cooling of electrons dominated by the synchrotron process, is efficient also at lower energies, resulting in a softer synchrotron spectrum.

\section{Example comparison with observations of Cyg~X-3}

Cyg~X-3 is the extreme case of the binary system with the microquasar features due to the presence of a very luminous WR type star (the radius
$R_\star = 2\times 10^{11}$ cm and the surface temperature $T_\star = 9\times 10^4$ K, see \citealp{cm94}).
It is also a very compact system (the separation of the components $D = 2.25R_\star$ \citealp{cm94}). 
It has been classified to belong to the microquasar class due to evidences of a relativistic motion. 
Recently, transient $\gamma$-ray emission has been detected from Cyg~X-3 before the major radio flare \citep{ab09a, tetal09}. 
The emission above 100 MeV was first modeled with a single power law with the differential spectral index $-2.7\pm 0.05_{stat}\pm 0.20_{syst}$ \citep{ab09a}. 
More recently, a curved power-law was found to describe the GeV data more precisely \citep{ab11}. 
The GeV emission shows modulation, of the order of $\sim 25\%$, with the maximum when the compact object is behind the companion star.
The level of this modulation seems to be much smaller than those observed in the TeV $\gamma$-ray emitting binaries such as LS 5039 \citep{ab09c} and LS I +61 303 \citep{ab09b}.
Note, that the baseline emission  in the Cyg~X-3 light curve may not be well known due to the strong background from the Galactic disk. 
However, if the reported level of modulation is real, then this might suggest a different mechanism of the modulation of the $\gamma$-ray signal seen in Cyg~X-3 from those ones observed in the TeV $\gamma$-ray binaries, e.g. precession of the disk/jet or partial contribution  to the $\gamma$-ray emission from the outer and inner parts of the jet.
The former hypothesis allows to explain the modulated and unmodulated $\gamma$-ray signal in terms of our model alone.
If the whole disk/jet system undergoes a precession in its movement around the centre of mass of the binary system, the angle at which we see the jet is constantly changing. 
As we showed above this will result in changing of the observed emission.
Even more interestingly, if this is the case we should observe the variability not only in the flux level, but also in the spectral shape (in particular in the cut-off energy).
Note moreover that such an effect would be easily distinguishable from the variations of the spectrum due to the absorption in the radiation field of the companion star. 
Due to large differences between the temperature of the disk and of the star those spectral features would occur at completely different energies. 
The latter hypothesis might be in accordance with our model in which  acceleration of electrons occurs already close to the base of the jet, resulting in the $\gamma$-ray production in the IC $e^\pm$ pair cascade occurring in the radiation field of the accretion disk.
 In this case, the low level modulation of the $\gamma$-ray signal might become a consequence of the cascade developing in the radiation field of the 
massive star as previously considered by e.g., \citep{bed10}. 

\begin{figure}
\includegraphics[scale=0.74, trim=0  17 0 0, clip]{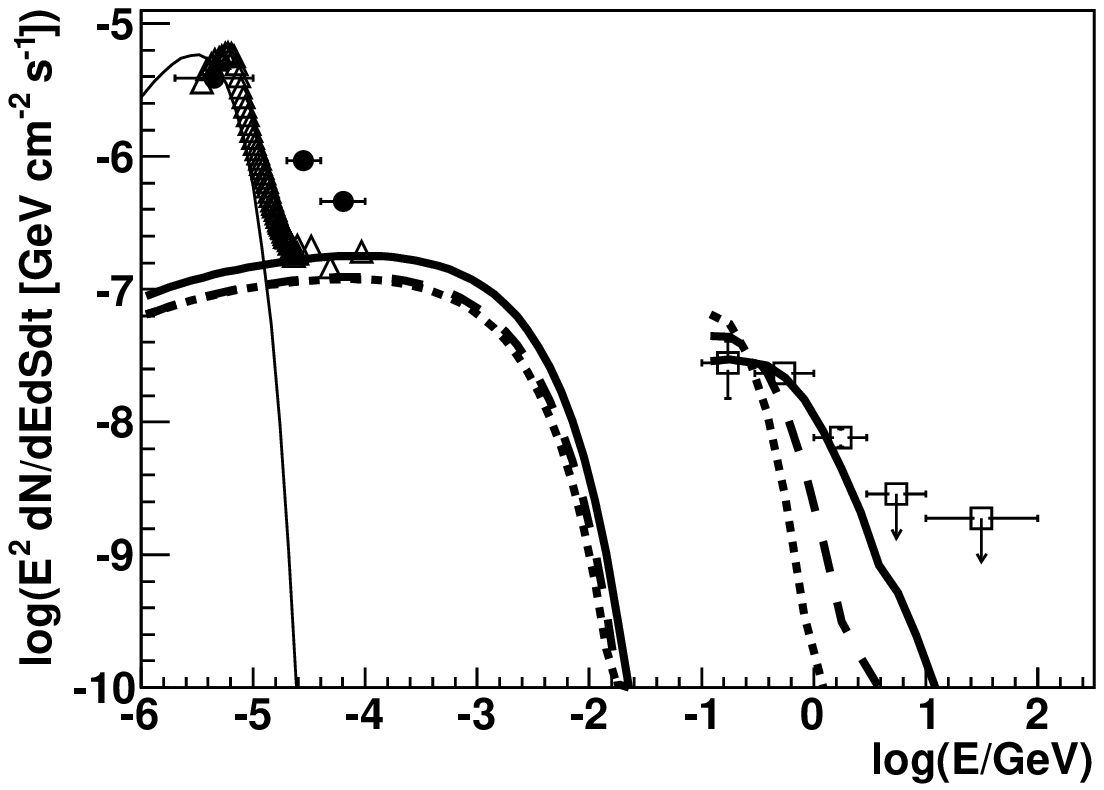} \\ %% 7.86244e+36 erg/s
\includegraphics[scale=0.74, trim=0  0 0 0, clip]{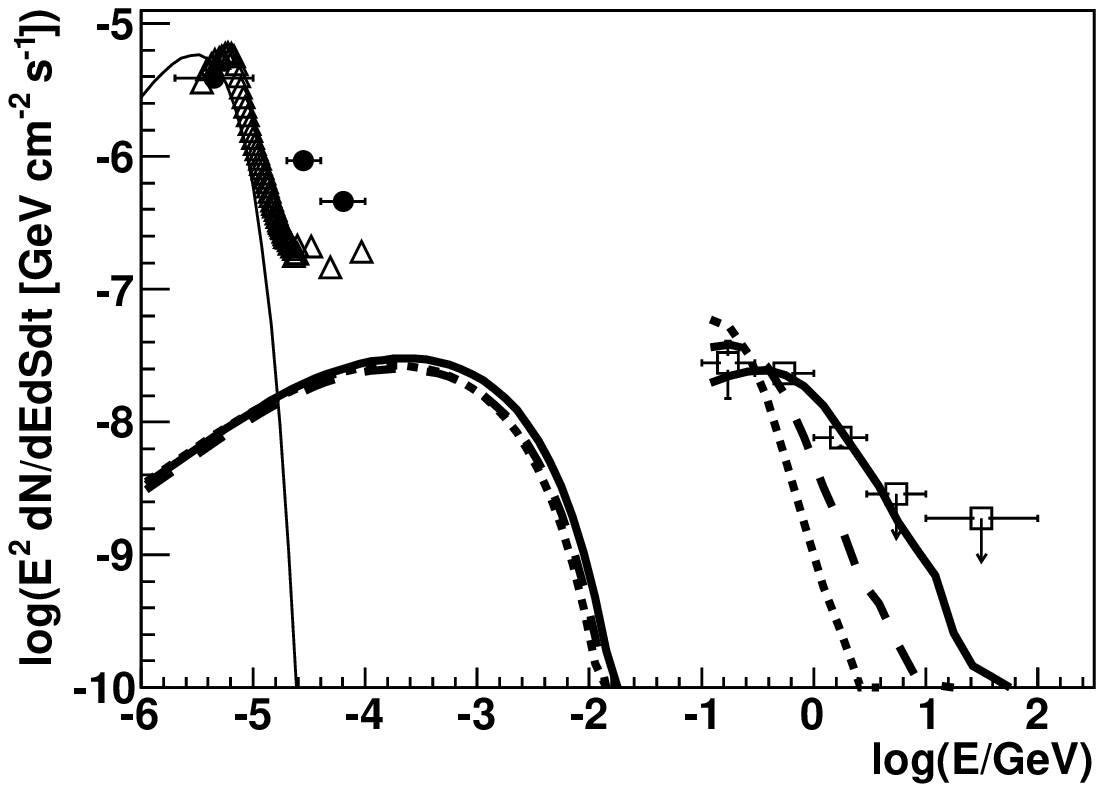} %% 1.84243e+36 erg/s
\caption{
SED produced in the IC $e^\pm$ pair cascade observed at different range of angles measured in respect to the jet axis:
$30-46^\circ$ (solid),
$46-59^\circ$ (dashed),
and $59-70^\circ$ (dotted),
for the disk parameters as expected from the Cyg~X-3 binary system ($\rin = 2\times10^6\,\mathrm{cm}$, $\tin = 1.5\times 10^7\,\mathrm{K}$).
The disk emission is represented with the thin solid curve. 
The results of various observations of Cyg~X-3 in the X-ray to $\gamma$-ray energies are shown with data points: 
BeppoSAX (the lowest energy full circle, \citealp{ve07}),
INTEGRAL (the other two full circles, \citealp{bi07})
RXTE (the empty triangles, \citealp{szm08}),
Fermi-LAT (the empty squares, \citealp{ab11}). 
Electrons are injected with the differential power law spectrum with an index $-1.75$ between 0.1~GeV and the maximum energy determined by balancing energy gains and losses (see Sect.~2.1) in a range of distances from the disk:  $H = 1-300\,\rin$.
The assumed velocity of the jet $v_{\rm b} = 0.5c$.
The acceleration coefficient $\xi=0.01$ and the equipartition parameter is $\eta=0.05$ (the top panel) and $10^{-3}$ (the bottom panel).
The spectra for specific ranges of the observation angles are normalized to the total flux seen by the Fermi-LAT telescope.} 
\label{fig_cygx3}
\end{figure}

We apply the likely parameters of the accretion disk around compact object in Cyg~X-3 ($\rin = 2\times10^6\,\mathrm{cm}$ and $\tin = 1.5\times 10^7\,\mathrm{K}$, consistent with the observed X-ray luminosity from the accretion disk, which is  of the order of $\sim 10^{38}$ erg s$^{-1}$).
The jet velocity is fixed on $v_{\rm b} = 0.5$c, within the range measured for Cyg~X-3 \citep{mio01, mil04}.
The inclination angle of the axis of the binary system (and probably also of the jet) to the observer in estimated in the range $i = 30^{\rm o} - 70^{\rm o}$ for the extreme parameters of the binary system (see \citealp{sz08}).
For these parameters we calculate the $\gamma$-ray spectra produced in the anisotropic IC $e^\pm$ pair cascade as considered above. 
We compare the spectra calculated in the framework of our model, with the Fermi-LAT observations of Cyg~X-3 assuming that bulk of its emission comes from the inner jet (see Fig.~\ref{fig_cygx3}).  

Investigation of the reasonable parameter range for our model shows that in order to reproduce the $\gamma$-ray spectrum observed by Fermi-LAT, the injection of electrons should occur up to at least a few hundred $\rin$. 
As an example, we show the specific injection case of electrons between $1-300\rin$ (see Fig.~11).
If this condition is not fulfilled, then strong absorption of $\gamma$-ray in the disk radiation cuts-off the GeV spectra at lower energies. 
Therefore, our modelling lower the restrictions presented by \citet{cer11}, who concluded that the emission region should be at least above $10-10^3\rin$ from the base of the jet.
The difference in our interpretation is due to the consideration of the full anisotropic cascade IC $e^\pm$ pair cascade in our case in contrary to the simple absorption arguments made by \citet{cer11}.   
The best agreement with the observational data is obtained for rather hard electron spectra, with the differential spectral index in the range 1.5 -- 2, and the injection power in electrons equal to $2-8 \times 10^{36}\,\mathrm{erg/s}$. 
Such a value is of the order of a few percent of the disk luminosity.  
According to our model, the low observation angle ($30-45^\circ$) is preferred. 
Interestingly, the flat part of the X-ray spectrum observed by RXTE above $\sim30\,\mathrm{keV}$ can be naturally explained in terms of our model as the synchrotron radiation of the primary electrons within the jet. 
Such an emission is strong enough for a rather large equipartition parameter ($\eta \sim 0.05$). 
For low values of $\eta$, the hard X-ray emission requires another active mechanism which can be located e.g. in the inner disk corona (e.g. \citealp{hja09}).

\section{Conclusions}

We considered for the first time the anisotropic IC $e^\pm$ pair cascade model for microquasars in which electrons injected in the inner part of the jet interact with the  radiation field of the accretion disk. 
In this model, the maximum energy to which the electrons are accelerated in the jet is determined by the balance of the energy gain, and the synchrotron and IC energy losses, with the additional condition on the ballistic escape time of the electrons being shorter then the acceleration time. We showed that the simplified model of a purely exponential absorption will result in neglecting  the $\gamma$-ray bump at energy of $\sim 1\,$GeV and a more steep spectrum in the multi-GeV energy range.
As a result, the constraints on the  location of the emission region within the jet should change in the full cascade scenario.
We considered two possible regimes for cooling of primary electrons, i.e.  dominated by either synchrotron or IC process. 
In the framework of those scenarios, we investigated the expected synchrotron and IC spectra and their dependence on the observation angle and location of the emission region.
We found out that for specific conditions, multi-GeV $\gamma$-ray emission can be produced and escape absorption from the radiation of the accretion disks in microquasars. 
Moreover, if the acceleration coefficient, $\xi$, is as large as $\sim 0.1$, even the TeV emission can be expected, providing that the jet is weakly magnetized. 
We suggest that such a scenario may explain the presence of some of unidentified TeV sources which do not show a clear radio nor X-ray counterpart. 

We also studied the effect of the possible transition of the inner part of the standard \citeauthor{ss73} accretion disk into a radiatively inefficient, advection dominated disk on the emerging spectra. 
We found out that the resulting spectra are significantly different in two ways. If the injection of the relativistic electrons occurs far from the disk, the $\gamma$-ray spectra observed at a large inclination angle are extending to higher energies due to the weaker absorption.
In addition, the $\gamma$-ray emission in the case of acceleration of the electrons close to the accretion disk is dimmer, due to lack of the strong radiation from the hot inner part of the disk. 

Finally, we also confront our model with the observations of the Cyg~X-3 microquasar performed in X-ray and GeV $\gamma$-ray energy ranges. 
We conclude that our model can explain the GeV $\gamma$-ray emission seen from this source, provided that electrons are accelerated within $\sim 300\,\rin$ from the base of the jet with a rather hard differential spectral slope of $-1.5$ to $-2$. Moreover relatively small observation angle is required ($30^\circ-45^\circ$).
While the X-ray emission of Cyg~X-3 is clearly dominated by the emission from the disk, a flat part of the SED seen by RXTE above $30\,\mathrm{keV}$ can be explained in the framework of our model as the synchrotron radiation of the primary electrons cooling in the jet, provided that the equipartition parameter $\eta$ is close to $\sim 0.05$.

\section*{Acknowledgments}
This work is supported by the Polish MNiSzW grant 745/N-HESS-MAGIC/2010/0.

%%%%%%%%%%%%%%%%%%%%%%%%%%%%%%%%%%%%%%%%%%%%

\end{document}